\documentstyle[11pt,aaspp4,flushrt,psfig]{article}

\slugcomment{To appear in {\sl ApJ}}

\def\etal{\hbox{\it et al.}$\,$}
\def\simlt{\lower.5ex\hbox{$\; \buildrel < \over \sim \;$}}
\def\simgt{\lower.5ex\hbox{$\; \buildrel > \over \sim \;$}}
\def\sext{{\sl SExtractor}}
\def\taurus2{{\sc Taurus-2}}
\def\lineunits{erg\,s$^{-1}$\,cm$^{-2}$}
\def\lineSB{erg\,s$^{-1}$\,cm$^{-2}$\,$\Box ''$}
\def\ergsPerAng{erg\,s$^{-1}$\,cm$^{-2}$\,\AA$^{-1}$}
\def\ergs2band{erg\,s$^{-1}$\,cm$^{-2}$\,band$^{-1}$}
\def\ergsPersec{erg\,s$^{-1}$}

\def\as{$''$}
\def\am{$'$}

\def\aspix{$''$\,pix$^{-1}$}
\def\d{$^{\circ}$}
\def\MpcPer3{Mpc$^{-3}$}
\def\Mpc3{Mpc$^{3}$}

\def\kmsMpc{km\,s$^{-1}$\,Mpc$^{-1}$}
\def\sqam{$\Box '$}

\def\sqdeg{\,$\Box^\circ$}

\def\kmPerSec{km\,s$^{-1}$}
\def\persqdeg{${\Box^\circ}^{-1}$}
\def\Msunyr{$M_\odot$\,yr$^{-1}$}

\def\OII{[{\ion{O}{2}}]}
\def\OIII{[{\ion{O}{3}}]} 
\def\NII{[{\ion{N}{2}}]}
\def\SII{[{\ion{S}{2}}]} 

\def\Ha{{H$\alpha$}\/} 
\def\Hb{{H$\beta$}\/}

\def\Sii{[{\ion{S}{2}}]~$\lambda\lambda$6717,6731\/}

\def\Nboth{[{\ion{N}{2}}]~$\lambda\lambda$6548,6583\/}

\lefthead{Jones \& Bland-Hawthorn}  \righthead{}

\begin{document}

\title{THE TAURUS TUNABLE FILTER FIELD GALAXY SURVEY: SAMPLE SELECTION
AND NARROWBAND NUMBER-COUNTS}

\author{D.\ Heath Jones\altaffilmark{1}}
\affil{Research School of Astronomy and Astrophysics, 
Australian National University, 
Mount Stromlo Observatory,
Cotter Road,
Weston ACT 2611, 
Australia}

\author{Joss Bland-Hawthorn}
\affil{Anglo-Australian Observatory, P.O.\ Box 296, Epping NSW 2121, Australia;
 jbh@aaoepp.aao.gov.au}

\altaffiltext{1}{{\em Current Address.}~~European Southern Observatory,
Casilla 19001, Santiago 19, Chile; hjones@eso.org}

\begin{abstract}
Recent evidence suggests a falling volume-averaged star-formation rate 
(SFR) over $z \simlt 1$. It is not clear, however, the extent to 
which the selection of such samples influences the measurement of 
this quantity. Using the {\sc Taurus} Tunable Filter (TTF) we have
obtained an emission-line sample of faint star-forming galaxies over
comparable lookback times: the {\sl TTF Field Galaxy Survey}.
By selecting through emission-lines, we are screening galaxies through a
quantity that scales directly with star-formation activity for a 
given choice of initial mass function. The scanning narrowband 
technique furnishes a galaxy sample that differs from 
traditional broadband-selected surveys in both its volume-limited 
nature and selection of galaxies through emission-line flux. 
Three discrete wavelength intervals are covered, centered at 
\Ha\ redshifts $z = 0.08$, 0.24 and 0.39.

Galaxy characteristics are presented and comparisons
made with existing surveys of both broadband and emission-line selection.
Little overlap is found in a direct comparison between the
{\sl TTF Field Galaxy Survey} and a traditional galaxy redshift survey, 
due to the respective volume and flux limitations of each.
When the number-counts of emission-line objects are compared with
those expected on the basis of existing \Ha\ surveys,
we find an excess of $\sim 3$ times at the faintest limits.
While these detections are yet to be independently confirmed, 
inspection of the stronger subsample of galaxies detected in
both the line and continuum (line-on-continuum subsample; 13\%)
is sufficient to support an excess population. The faintest objects
are galaxies with little or no continuum, rendering them undetectable
by conventional redshift surveys. This increase in the emission-line
field population implies higher star-formation densities over 
$z \simlt 0.4$. However, further study in the form of multi-object 
spectroscopic follow-up is necessary to quantify this and confirm the
faintest detections in the sample.
\end{abstract}



\section{INTRODUCTION}

The decline in average star-formation rate provides a fundamental
constraint to hierarchical models of galaxy formation, over half
the age of the universe. 
Using the {\sc Taurus} Tunable Filter (TTF) on the Anglo-Australian
Telescope (AAT) we have obtained an emission-line sample of faint
star-forming galaxies over half the age of the universe.
This {\sl TTF Field Galaxy Survey} (\cite{jones99}~1999) is 
primarily sensitive to star-forming galaxies seen through \Ha\ at three
discrete wavelength intervals, corresponding to redshifts 
$z = 0.08$, 0.24 and 0.39.
In selecting through emission-lines, we are using a quantity that
scales directly with star-formation rate, effectively
selecting galaxies by precisely the parameter we aim to measure.
By scanning with wavelength, we automatically define a strict volume-limited
sample, less affected by the magnitude-volume sampling effects of conventional
broadband surveys. Thus, the tunable filter technique provides a unique
means of quantifying the star-formation activity across redshifts 
relevant to interpretation of the most recent star-formation
history of the universe. 

The red {\sc Taurus} Tunable Filter  
is a tunable Fabry-Perot Interferometer 
covering 6500$-$9600~\AA\ (\cite{blajon98}~1998a,b). 
The spacing between the TTF plates is much narrower 
and adjustable over a wider range than conventional etalons, 
thereby giving its function as a tunable narrow passband. 
In effect, the TTF affords monochromatic imaging 
with an adjustable passband of between 6 and 60~\AA.
TTF has an important advantage in the detection of
faint line emission over conventional redshift surveys. Deep
pencil-beam surveys typically pre-select objects down to $B \sim 24$ 
(\cite{glazebrook95}~1995) or $I \sim 22$ 
(\cite{lilly95}~1995), since these represent the 
practical spectroscopic limits on the 4~meter telescopes used.
However, these limits are set assuming the objects are dominated
by continuum. Objects with continuum flux beyond the spectroscopic
limit but with emission-lines above the limit could
be detected by these same instruments, albeit through the lines alone.
However, because the initial selection is made from the
broadband flux, such objects are excluded {\sl at the outset} on 
the basis of having too faint a continuum level.

In order to match TTF, a broadband-selected redshift survey needs
to select objects to a fainter limit, in the knowledge that
for the faintest objects, only those with emission-lines will
register a detection spectroscopically. However, this
represents a vastly inefficient way of finding emission-line galaxies.
Between $B \sim 24$ and $B \sim 28$, the number of objects on the
sky (per magnitude interval) increases by an order of 
magnitude (\cite{metcalfe95}~1995).
Such object densities would be impractical to search given the
limitations of current multi-slit spectrographs in terms of
object multiplex. In this paper we compare the very
different emission-line galaxy samples obtained by the 
{\sl TTF Field Galaxy Survey} and the Autofib Galaxy Redshift
Survey (\cite{ellis96}~1996) in regions where the two overlap.

In this paper we describe the {\sl TTF Field Galaxy Survey}.
Section \ref{s:strategy} discusses the adopted survey strategy and
the reasons for this approach. Section \ref{s:selection} describes the
selection of the main emission-line sample and its attributes.
Known selection effects and potential sources of contamination are
reviewed and their effect on the sample discussed.
The main characteristics of the sample measurable from 
these TTF data are given in Sect.~\ref{s:character}. Included in this 
section are narrowband number-counts and a comparison of our 
narrowband selected emission-line galaxies with those of the 
broadband-selected Autofib Galaxy Redshift Survey.
We also derive preliminary \Ha\ luminosity functions
and compare them to those of existing surveys.
Concluding remarks are made in Sect.~\ref{s:conclremarks}.
Unless stated otherwise, we assume a Friedmann model cosmology
(cosmological constant $\Lambda_0 = 0$) with a Hubble constant
$H_0 = 50$ \kmsMpc\ and deceleration parameter $q_0 = 0.5$.


\section{SURVEY STRATEGY\label{s:strategy}}

\subsection{Survey Coverage \label{s:surveyCoverage}}

Considerable freedom exists in a scanning narrowband survey conducted 
with a tunable filter, by virtue of the flexibility of the instrument. 
The elements that remain fixed are field coverage on the
sky and the Airy-profile shape of the passband. The adjustable parameters 
are the passband width, the passband coverage and
sampling rate with wavelength, and the exposure time per slice.
Individually, each affects the number of objects obtained and a 
judicious selection is necessary if sample size is to be maximized.

The {\sl TTF Field Galaxy Survey} is a survey for redshifted
emission-line galaxies in the field. All fields were taken with TTF 
in \taurus2\ at the $f/8$ Cassegrain focus of the Anglo-Australian
Telescope (AAT).
The survey comprises 15 scans of 10
slices, at random high-galactic latitude fields scattered around the sky.
By {\sl slice}, we mean an exposure at a particular wavelength, or
the image obtained by co-adding many such exposures.
The total sky coverage is 0.27~\sqdeg\ with most fields selected on
the celestial equator for maximum accessibility from both 
northern and southern observatories.
The scans are distributed between the 707/26 ($R_1$), 814/33 ($R_5$) 
and 909/40 ($R_8$) TTF blocking filters. Each scan was repeated 3 times
with the telescope offset by $\sim 10$\as\ between each so that cosmic-rays
and ghost images could be removed through median filtering. Two CCDs
were used: a $4096 \times 2048$~pix MIT-LL and a $1024 \times 1024$~pix
Tektronix with pixel scales of 0.371 and 0.594~\aspix\ respectively 
at AAT $f/8$. In both cases,
individual fields were limited by a circular field stop of 9\am\
diameter.

Table~\ref{t:stepping} summarizes the passband settings as used for scans 
in each spectral region.
Mean passbands with FWHM ($\delta \lambda$) of 12.9, 16.4 
and 22.3~\AA\  were used to cover
707/26, 814/33 and 909/40 respectively. In velocity space the passbands
had widths of 548, 604 and 725~\kmPerSec\ respectively. 
We would have preferred to maintain a fixed velocity width
between wavelength windows although this proved difficult in practice
since many of the TTF spectral regions were being wavelength-calibrated 
for the very first time during these observations.
Passbands were stepped in increments
of $1.3\,\delta\lambda$ as a compromise between velocity
coverage and sampling continuity. 

Table \ref{t:scans} gives the observational characteristics
for scans in each of the three spectral regions, in terms of \Ha\ and \OII.
For \Ha, the spectral regions sample redshift windows at
$z = 0.08$, $0.24$ and $0.39$. Note that the passbands used include
the contribution of nearby \NII\ emission with \Ha.
Corresponding redshifts in the \OII\
line are $z = 0.90$, $1.18$ and $1.47$, although we expect fewer \OII\
galaxies at our flux limits. We attempted to compensate somewhat
for the increased volume (per scan) represented at higher redshifts
and longer wavelengths. As such, 7 fields were scanned in the
707/26 filter, 5 in the 814/33 and 3 in the 909/40. This arrangement
provided approximately equal volumes in \OII\ ($\sim 48000$ \Mpc3)
but a large range in \Ha\ ($1000$ -- $10000$ \Mpc3), because of the low
\Ha\ redshift for the 707/26 window. The only way to match the low and
high redshift volumes in an efficient manner would be through a 
wide-field instrument such as an objective prism on a Schmidt telescope.
However, these are typically 6 magnitudes less-sensitive than
TTF in the case of photographic surveys (e.g.~\cite{zamorano94}~1994)
and 4 magnitudes with the use of a CCD (\cite{salzer00}~2000).
For wide-field tunable filter searches, 
the main limitation is the degradation of bandpass 
transmission that an interference device such as a tunable filter experiences
in the fast beam of a wide-field instrument (\cite{lissberger59}~1959).
\cite{bland01}~(2001) describe the design of a wide-field tunable Lyot filter 
that avoids these difficulties.

Exposure times were typically 600~s per slice for the 707/26 and 814/33 scans
and 1080~s per slice for the 909/40 scans. Of the 13 nights obtained
on the Anglo-Australian Telescope (AAT) during 1997 April -- 1998 April,
7 were photometric with reliable flux measurements. Only scans taken under
photometric conditions have been included. These exposure limits
permit us to reach $0.5 \times 10^{-16}$~\ergs2band\ as a 
$3\sigma$-detection in a 2\as\ aperture, in our deepest exposures
for the 909/40 interval. For a 20~\AA\
bandpass, this flux is equivalent to $2.5 \times 10^{-18}$ \ergsPerAng\ or
$I = 21.8$ as broadband continuum. Alternatively, if the
$0.5 \times 10^{-16}$~\ergs2band\ is confined solely to a {\sl line},
(with no other continuum), then this is equivalent to a broadband
detection limit of $I = 26.8$.
The instantaneous star-formation rate (SFR) scales directly with
the \Ha\ and \OII\ line luminosity. Using the calibrations for \Ha\ and
\OII\ from the literature review of \cite{kennicutt98}~(1998)
we can determine the limiting star-formation rates of our sample.
A limiting flux detection of $\sim 0.5 \times 10^{-16}$~\ergs2band\
is sufficient to detect and measure star-formation rates of a few
tenths \Msunyr, comparable with SMC-type levels at the nearest redshifts
and the LMC at $z_{{\rm H}\alpha} \sim 0.4$.

The narrowband scanning of TTF selects a galaxy sample that differs
in two important ways from conventional redshift surveys.
First, the narrowband technique produces
a {\sl volume-limited} sample of emission-line galaxies. The volume
of redshift space covered is chosen by the observer through the range
and placement of narrowbands. This differs from the broadband limit of
conventional redshift surveys which allows galaxies from a wide range
in redshift, making total volume a complex function of redshift.
The second difference is that TTF selects objects through narrowband
flux which is sensitive to the line flux and therefore star-formation
rate. Most conventional redshift surveys define emission-line sub-samples
from the initial broadband-selected sample. 
The inclusion of a galaxy in such samples is dictated by the continuum of the
combined stellar population and not by the instantaneous star-formation rate. 
To define a complete star-forming sample over a given volume, a narrow 
bandpass well-matched to flux detection in an emission-line is essential.

Since the TTF technique is without precedent we directed 9 of our fields
to include emission-line galaxies found as part of the Autofib galaxy redshift
catalogue (\cite{ellis96}~1996). The Autofib catalogue contains over 
1700 galaxies in the apparent magnitude range $11.5 < b_J < 24.0$. 
Since it is the combination of many optical surveys, emission-line 
galaxies are detected primarily through \OII. Of the TTF fields observed, 
there are 18 emission-line galaxies in the Autofib catalogue with redshifts 
placing \Ha\ in one of either the 707/26, 814/33 or 909/40 blocking filters.
We defined the Autofib emission-line sample as those galaxies with
\OII\ rest-frame equivalent widths (REWs) greater than 10 \AA.
Through the galaxies common to both catalogues,
we can examine how representative a broadband-selected survey is of the
underlying star-forming population. Furthermore, we can gain insight
into how the TTF detection rate goes as a function of broadband magnitude.

Table \ref{t:observations} shows the observations undertaken
for the {\sl TTF Field Galaxy
Survey}. The number of Autofib galaxies expected in \Ha\ is also included.
Figure \ref{figaitoff} shows the distribution of fields on the sky. The fields
overlapping with the Autofib galaxies are indicated.
We made a search of the NASA Extragalactic Database\footnote{
Operated by the Jet Propulsion Laboratory, California Institute of
Technology, under contract with the National Aeronautics and Space
Administration.}
(NED) to check
whether any known clusters lie near our fields. No clusters
with measured redshifts overlap with our fields although
4 of the optically-identified cluster candidates of \cite{lidman96}~(1996)
lie within 6\am\ radius of our 10\_3F, 13\_3E, 10\_3C and 10\_3H
field centres. Within the \cite{lidman96} sample, these cluster
candidates are all rated with low significance. 
However, we later examine the number density and wavelength distribution of
emission-line candidates within these fields as a check.

\subsection{Background Source Contamination\label{s:volLimited}}

All narrowband imaging emission-line searches are open to different lines
at varying redshift. Any line for which both the rest wavelength and 
source redshift act to place it within a TTF survey interval will be detected
if sufficiently luminous. The fixed flux detection limit translates
to different limiting luminosities in each emission-line, depending 
on its redshift. Furthermore, the fixed wavelength coverage of the scan 
defines different co-moving volumes at each of the redshifts. 
This needs to be taken into account
when optimising survey coverage, not just in terms of object density,
but also in the {\sl type} of emission-lines sought, since some
are preferred to others for measuring star-formation rates
(\cite{kennicutt98}~1992a, 1998).

In a blind search at these wavelengths the emission-lines we would most
expect to see in galaxies are \Ha, \OII, \Hb\ and \OIII\ 
(\cite{tresse99}~1999; \cite{folkes99} 1999; \cite{kennicutt92b}~1992b). 
The TTF passband is too wide to
separate \Ha\ and at least one of \Nboth. 
The detection of Ly$\alpha$ would be rare at
the levels we are probing. 
Recent estimates for the density of 
Ly$\alpha$ emission-line galaxies with line fluxes of 3 to 
$5 \times 10^{-17}$~\lineunits\ number $\sim 4000$~\persqdeg\ per
unit $z$ (\cite{rhoads00}~2000). This implies negligible numbers of 
Ly$\alpha$-emitters at our brighter flux limit 
($\sim 1 \times 10^{-16}$~\lineunits).
\Sii\ is another line we do not expect to find since it is are too 
far into the red to be found in an appreciable volume. 
In any narrowband survey, we expect a competing luminosity-volume effect 
between \Ha\ being probed to fainter luminosities over smaller volumes, and
background emission-lines at brighter luminosities but larger volumes.

\Ha\ is the most preferred line for measuring the instantaneous 
star-formation rate, being the strongest Balmer recombination line and
in the part of the optical spectrum affected least by extinction 
(\cite{kennicutt92a}~1992a;  \cite{vdWerf97}~1997; \cite{tresmad98}~1998; 
\cite{glazebrook99}~1999). It is directly related to the number of short-lived
OB stars, indicative of the stellar birthrate in a galaxy. These hot, massive
stars produce ultraviolet (UV) radiation that ionises the nearby gas to 
create an \ion{H}{2} region. Recombination produces the observed 
emission-lines and of these, \Ha\ most closely traces
the amount of ionising UV flux. Of the other Balmer lines, \Hb\ is
weaker and prone to filling-in by the underlying stellar absorption.
\OII\ can be also be related to the instantaneous star-formation rate,
although the underlying assumptions are less robust, 
since this (and other lines such as \NII, \OIII\ and \SII), 
depend heavily on the metal fraction of the gas 
(\cite{kennicutt92a}~1992a). In the case of lines with a higher
ionisation potential, the energy of the incident UV flux is also
an important factor. As a consequence, they are
unreliable star-formation indicators. Of all the optical lines,
\Ha\ is the preferred choice. 

Given the reasons outlined above, \Ha\ is the obvious target for our survey
of star-formation. However, it is prudent for us to examine the numbers
of background emission-line galaxies likely from a scanning narrowband survey.
Accordingly we made estimates of 
the galaxy numbers expected from each of the other major optical
emission lines relative to \Ha. For a Friedmann model cosmology
with $\Lambda_0 = 0$, the co-moving volume of space sampled
in an emission-line at redshift $z$, is
\begin{equation}
\Delta V = \frac{4 \pi\, d_{\rm M}^2 \, c \, \Delta z \, A_{\rm field}}
{1.4851 \times 10^8 \cdot H_0 (1+z) (1+ \Omega_0 z)^{1/2}} .
\label{dV}
\end{equation}
Here, $A_{\rm field}$ is the area of the field on the $1.4851 \times
10^8$ \sqam\ of the sky and $\Delta z$ is the redshift coverage of the scan
and $d_{\rm M}$ is the proper distance.
Galaxy emission-line luminosity functions
have been determined by several authors for H$\alpha$ at a range
of redshift (\cite{gallego95}~1995; \cite{tresmad98}~1998; 
\cite{gronwall98}~1998; \cite{yan99}~1999).
In most of these cases, the \Ha\ luminosity function is found to be
well-matched by a \cite{schechter76}~(1976) function,
\begin{equation}
\phi(L)\, dL = \phi^* \, (L/L^*)^\alpha \, \exp(-L/L^*) \, d(L/L^*) ,
\label{schecht}
\end{equation}
where $\phi(L)$ is the number of galaxies per unit volume
at luminosity $L$. The parameters $\phi^*$, $L^*$ and $\alpha$
are constants describing the shape of the distribution.
More specifically, the density of H$\alpha$ galaxies per unit volume, 
$\Phi[\log(L_{{\rm H}\alpha})]$, is binned per
0.4 interval in $\log(L_{{\rm H}\alpha})$, and can be related
to the Schechter description through
\begin{equation}
\Phi \biggl [ \log(L_{{\rm H}\alpha}) \biggr ] \, 
 \frac{d \, \log L_{{\rm H}\alpha}}{0.4} =
\phi(L_{{\rm H}\alpha}) \, d L_{{\rm H}\alpha},
\label{transfm}
\end{equation}
for $L = L_{{\rm H}\alpha}$, luminosity in H$\alpha$ (\cite{gallego95}~1995).
The total number $N(L_{\rm lim})$ 
of \Ha\ emission-line galaxies (per unit volume)
detected to limiting luminosity $L_{\rm lim}$
can be obtained by integrating Eqn.~(\ref{schecht}) to that luminosity,
\begin{equation}
\begin{array}{rl}
N(L_{\rm lim}) & = \int_{L_{\rm lim}}^{\infty} \phi(L_{{\rm H}\alpha})\,d L_{{\rm H}\alpha} \\
   &   \\
   &  = \phi^* \, \Gamma(\alpha+1) \cdot \bigl [1- P(\alpha + 1, L_{\rm lim}/L^* ) \bigr ] . \\
\end{array}
\label{totNumber}
\end{equation}
Here, $P(\alpha + 1, L_{\rm lim}/L^* )$ is the incomplete gamma 
function\footnote{
By definition, the incomplete gamma function is
$P(a,x) = \int_0^x e^{-t} t^{a-1} \, dt / \Gamma(a)$, where $\Gamma(a)$
is the gamma function and $a > 0$.
}
evaluated in terms of the Schechter parameters.
We applied the function in Eqn.~(\ref{totNumber}) to estimate the
total numbers of \Ha\ emission-line galaxies seen in each of the
TTF scan intervals. As the \Ha\ luminosity function evolves rapidly
with redshift, we assumed an evolving $\phi^*$, $L^*$ and $\alpha$
of the forms
\begin{equation}
\begin{array}{c}
\phi^* (z) = \phi_0^* \, (1+z)^{\gamma_\phi}, \\
L^* (z) = L_0^* \, (1+z)^{\gamma_L} , \\
\alpha(z) = \alpha_0 \, + \gamma_\alpha z , \\
\end{array}
\label{phiLwithZ}
\end{equation}
following the forms adopted by \cite{heyl97}~(1997) for evolution
in the broadband luminosity function.
The free parameters $\phi_0^*$, $L_0^*$, $\alpha_0$, 
$\gamma_\phi$, $\gamma_L$ and $\gamma_\alpha$
were constrained by the luminosity functions 
of \cite{gallego95}~(1995) at $z = 0$ 
($\phi^* = 0.63 \times 10^{-3}$ \MpcPer3, $L^* = 10^{42.15}$~\ergsPersec,
$\alpha = -1.3$) and
\cite{tresmad98}~(1998) at $z= 0.2$
($\phi^* = 1.48 \times 10^{-3}$ \MpcPer3, $L^* = 10^{42.13}$~\ergsPersec,
$\alpha = -1.35$). Table~\ref{t:calculations} lists the parameters describing
describing our adopted \Ha\ luminosity function evolution, as constrained
by these two samples.

With no equivalent emission-line luminosity functions for lines of
\OII, \Hb\ and \OIII, our best approach is to scale the \Ha\ 
luminosities by the mean line flux ratios of \cite{kennicutt92a}~(1992a).
Kennicutt deliberately selected his sample of E-Irr galaxies to match the
mix of morphological types seen in the Revised Shapley-Ames (RSA) Catalog
(\cite{sandage81}~1981). As such, it can not be considered to be
representative of the true space density these types.\footnote{
For example, the RSA Catalog under-represents the true space density
of Sd and later types due to their fainter mean absolute magnitudes
compared to early-type galaxies (\cite{sandage81}~1981).
}
Accordingly, we computed
mean flux ratios for \OII, \Hb, \OIII, \SII\ and \NII\, weighting
\cite{kennicutt92a}'s galaxy types by their occurrence. Measurements of 
$\phi^*$ by \cite{heyl97}~(1997) were used to estimate the relative space
densities of different types and their evolution with redshift. 
Figure~\ref{figkenevol} shows that under such a weighting
scheme, all line ratios show little change out to $z \sim 1$.
This is consistent with \cite{glazebrook99}~(1999) who find 
broad agreement between the \Ha/\OII\ ratios of their $z \sim 1 $ 
sample and the local \cite{kennicutt92a}~(1992a) sample.
The greatest change is the sharp decrease in \NII/\Ha\ ratios 
over $z \simlt 0.2$. We suspect this is due to the steep decline
in the numbers of early-type galaxies (Fig.~\ref{figkenevol},
{\em inset}),
in which strong \NII\ emission is most prevalent
(\cite{phillips86}~1986).
This constancy in line flux ratios is indeed what we might expect, given
a field population composed almost exclusively of
star-forming galaxies at such lookback times
(e.g.~\cite{ellis96}~1996; \cite{tresse99}~1999).
The effects of extinction aside, line ratios will only 
exhibit significant variation in cases of
active galactic nuclei (\cite{kennicutt92a}~1992a),
for which \cite{sarajedini99}~(1999) find no density evolution over 
$z < 0.4$ and mild evolution between $z=0.4$ and $z=0.8$.

Figure~\ref{figcontamEVOL} shows the relative predicted numbers that
our calculations yield for each of the 707/26, 814/33 and 909/40
TTF spectral windows. Numbers are shown as a function of flux for
both individual flux bins (Fig.~\ref{figcontamEVOL}, {\em left}) and
cumulatively with decreasing flux (Fig.~\ref{figcontamEVOL}, {\em right}).
We have assumed the galaxies to have a mean internal extinction of
1~mag at \Ha\ and a mean \Ha/(\Ha$+$\NII) flux ratio of 0.69
(\cite{tresmad98}~1998). Note however that both \cite{tresmad98} and
\cite{tresse99}~(1999) find a trend in \NII\ flux against \Ha.
We have not included this in the correction, since both surveys
use slit spectroscopy (1.75\as\ and 8\as\ slits respectively)
which misses light from the outer regions of large galaxies in a 
way that narrowband imaging does not.

The flux limits of our raw sample are $\sim 0.5$ to 
$1 \times 10^{-16}$~\lineunits.
As expected, \Ha\ galaxies dominate the cumulative numbers
at the brightest fluxes because
background galaxies in the other lines need to be intrinsically
more luminous, and hence, much rarer objects. At the fainter fluxes,
we find a broader mix of emission-lines as the larger volumes sampled for the
background lines begin to balance the fainter luminosity limits
probed in \Ha. 

Around the flux limits of our survey, \Ha\ dominates over \OII\, 
thereby representing more than 90\%\ of the full emission-line sample
in the redder 814/33 and 909/40 bands. In the 707/26 band, the \Ha\ 
and \OII\ galaxies comprise approximately equal numbers ($\sim 50$\%)
with comparatively fewer numbers of galaxies seen in \OIII\ and \Hb.
The dominance of \Ha\ at stronger fluxes in all bands suggests that
samples restricted to these values will essentially constitute
\Ha, with little contamination from background \OII. In the case of
707/26, this means truncating the emission-line sample artificially
higher ($\sim 1.5$ to $2 \times 10^{-16}$~\lineunits) than the 
natural flux limits of the survey. The trade-off is that there are
far fewer galaxies with higher line fluxes.
In short, we expect to \Ha\ galaxies to dominate samples
at fluxes exceeding $\sim 2 \times 10^{-16}$~\lineunits\ for the 707
interval, and $\sim 0.5 \times 10^{-16}$~\lineunits\ for the 814 and 909 
intervals.


\section{EMISSION-LINE SAMPLE\label{s:selection}}

\subsection{Candidate Selection}

The observations were reduced as follows. 
Individual frames were bias-subtracted and divided
by flatfields. Night-sky rings were removed and images
aligned and co-added in three separate scans. Object detection and
measurement was undertaken using \sext\ (\cite{bert96}~1996). Objects were
detected from the original frames and subsequently photometered from a
matching set of frames degraded to match the worst seeing frame.
Cosmic-ray and ghost image catalogues were constructed for each scan and
used to remove these detections from the raw emission-line 
candidate catalogues.
Emission-line candidates in which line flux was detected and continuum was 
not were selected by the line detections appearing in one or two 
adjacent frames and no others. These we term our {\sl line-only} candidates.
Candidates showing line flux on an otherwise detectable continuum 
we call our {\sl line-on-continuum} candidates. These
were found through a more complex procedure. Initially, all objects
in the field with continuum detected in 5 or more frames had a
straight-line of best-fit iteratively fit to the continuum
flux with wavelength. Fluxes deviant from this fit were rejected on
successive iterations refining the fit. Any objects
showing flux deviations in one or two adjacent frames above the
flux uncertainties of the continuum were included in the emission-line
catalogue. All objects in this group (in excess of $\sim 200$) 
were inspected on a mosaic of postage-stamp images (through the scan)
to confirm the rejection of cosmic-rays and ghosts.
The quality of the deviation above the fit was also checked to confirm
the fit as true to the continuum and unaffected by bad data points.
The continuum and emission-line fluxes in both sets were calibrated
through standard star observations from the same nights. Measured
fluxes were corrected for extinction by our own galaxy (albeit small since
the fields were all at high galactic latitude) and for flux truncation
by the apertures where necessary. 
The specific procedures are described by \cite{jones99}~(1999).

Figure~\ref{d13e7Best} shows example scans and TTF spectra for a
subset of candidates from one of the {\sl TTF Field Galaxy Survey} fields.
All of the objects shown are line-on-continuum candidates, for which
the continuum flux has been fit in order to find the emission.
Preliminary and final continuum fits are shown in the spectra
({\em dotted} and {\em dashed} lines respectively). One object
(214.14) is one of the 18 Autofib galaxies within our fields.
In detection terms, the relative strength of an emission-line
is quantified in terms of its {\sl $\sigma$-deviation}. This is the
line flux in units of background $\sigma$. For line-only candidates, $\sigma$
is the background RMS of the local sky noise against which the detection
was made. For example, a line-only detection with a $\sigma$-deviation 
of 3 is a 3$\sigma$-detection in the usual sense.
For line-on-continuum candidates, the situation is complicated by the
presence of continuum. In this case, the ``background $\sigma$'' is
that determined by either the RMS scatter of fluxes about the
continuum fit, or the mean uncertainty in the continuum fluxes,
whichever is greatest. Hence, objects with strong emission flux might
be rated with a low $\sigma$-deviation if that emission is superposed
on a continuum of comparable or greater strength. The $\sigma$-deviation is
an expression of the {\sl ease} with which an emission-line was detected,
not its strength in flux.

The raw catalogue of emission-line candidates was selected
from all objects with $\sigma$-deviations of 3 or greater.
At $3\sigma$, those candidates seen as a peak in a
single frame ideally require confirmation through a follow-up observation
such as spectroscopy.  
Figure~\ref{figlimits_sig} shows the distribution of $\sigma$-deviations
for each field. The line-only detections ({\em horizontal tickmarks}) and 
line-on-continuum detections ({\em crosses}) are shown separately. 
Observe in Figs.~\ref{figlimits_sig}($b$) and ($c$) that from the point
of view of detection, the majority of candidates ($\simgt 50$\%) are
weak detections of between 3 and 4$\sigma$, as might be expected.
Most fields show
a similar distribution of $\sigma$-deviations with respect to each other.
The exception is the 00\_3A field which shows substantial numbers
of strong $\sigma$-deviation objects in both cases of object.
We suspect that this may be due to either the
field being observed under above-average conditions and/or a possible
chance alignment with a galaxy cluster. 
We also note
that this field has 
comparatively higher numbers of faint line-only detections,
consistent with either hypothesis.

Finally, note that 
that there is a systematic offset between the limiting
line flux of the line-only sample and its corresponding line-on-continuum
sample for the same field. This is simply a reflection of the fact that a 
$3\sigma$-deviation detection seen against background sky can probe
to fainter line levels than an equivalent detection measured against
object continuum. These differences in sensitivity under different
circumstances must be taken into account when deriving a 
homogeneous sub-sample of objects.

\subsection{Selection of a Homogeneous Sub-Sample\label{s:defHomog}}

The selection of a homogeneous sub-sample of emission-line candidates
across many fields must necessarily take into account several
observables simultaneously. There are
many factors (both external and intrinsic to an object)
that influence its detection probability. By {\sl homogeneous
sub-sample}, we mean a sample of emission-line objects that
conforms to a set of well-defined selection criteria, within which
we know our sample is complete, or near complete. By {\sl complete}, we mean a
100\% detection probability for all objects within a given set of selection
criteria. The detection probability can be assessed by generating artificial
datasets spanning similar object characteristics as the real sample.
Since the observational properties of the artificial objects
are known from the outset, a success rate for detection can be measured
when these objects are passed through the same selection methods as were
used for the real objects. This can be used to assess the amount of
incompleteness as a function of selection level and make for its
correction. In practice, selection cuts are placed at some threshold level, 
below which, the incompleteness corrections become
intolerably large. Corrections for incompleteness are made
by weighting object numbers by the known missing fraction, provided
the observed galaxy numbers are sufficiently large and the correction
factors small.

In the {\sl TTF Field Galaxy Survey}, all objects are first selected
on the basis of a detection by \sext, which becomes increasingly
incomplete at the faintest limits. The line-only candidates are subject
to this alone whereas the line-on-continuum objects undergo further
selection on the basis of line flux relative to continuum. The
detection probability of both procedures must be measured before
we can set bounds to define a homogeneous sub-sample. The bounds
should be such that we can combine both types of object into
a single uniform catalogue, within which the differences in selection
method are no longer important.
We examined the detection probabilities for the fields observed in
cases of worst seeing and/or highest sky background.
Images were constructed containing artificial objects under identical
conditions of seeing and background RMS as the real observations. 
The artificial objects populated a 
$10 \times 10$ grid of object sizes and line fluxes across a range of values 
encompassing those from the real data. Objects of a given size and flux
were duplicated 20 times.
\sext\ was executed on the artificial images using an identical
detection configuration as the real field concerned. 
Its success at finding
some fraction of objects at a given size and flux is a 
measure of the detection probability as a function of both parameters,
under the particular observing conditions. In all fields, the detection
probability fell away quite steeply from 100\% detectability, over
a relatively narrow strip along the grid.
Two fields (22\_3A for 707; 10\_3C for 909) showed
detection probabilities declining at higher flux levels than the
other fields in that group. These two fields also contribute
significantly less to the final object numbers.
Given this, and the high levels of
incompleteness they would impose on the other fields, we excluded
them from the homogeneous sub-samples.

Figure~\ref{figlimits5} shows just the raw line-only detection candidates
in terms of both mean detected flux and object size. The mean detected flux
for these objects is simply the line flux detected on a single
narrowband frame. The samples in Fig.~\ref{figlimits5} are
labeled by blocking filter and the 22\_3A (707) and 10\_3C (909)
candidates have been removed. The solid lines are contours of
detection probability for the {\sl single worst field}
of each panel, with the 60\%-contour emboldened. 
All other fields have contours at fainter flux levels, not shown on 
this figure. The significant number of points in the
$\simlt 20$\% region come from these more sensitive fields,
which exhibit completeness to fainter flux levels.
The detection contours show that the natural detection limit of a 
narrowband survey such as this is emission-line surface brightness.
Accordingly, we apply two cuts: a minimum line flux cut and
a minimum surface brightness cut. Taken together, they select 
candidates from {\sl all} fields from within the 60\% detection 
contour of the {\sl worst} field. Individually, all fields except
the worst are complete to the selection limits imposed by the latter.
Note that because this contour only applies to the worst field,
Table~\ref{t:selectlimits} lists the flux and surface
brightness cuts. A value for surface brightness at the object center is
inferred from its FWHM and total flux, assuming an exponential disk
profile.

In addition to the cuts above, the line-on-continuum objects are subject to 
further selection based on the strength of the line flux {\sl relative}
to the continuum. This is necessary because of the influence that
the continuum level can have in determining whether an emission-line
is detected, irrespective of line flux. Figure~\ref{figlimits6}
demonstrates how we apply these cuts to the line-on-continuum
objects. Initially, we apply identical flux and surface brightness
selection for the line-only candidates, since both
catalogues depend upon object detection in the first instance
(Fig.~\ref{figlimits6}, {\em left}). Note that in this case,
the mean detected flux is the object {\sl continuum}, not the line
as in the line-only case. Second, we restrict the line fluxes
to the {\sl same} level as the mean detected continuum flux
(Fig.~\ref{figlimits6}~{\em right}, vertical line). This is
because the same detected flux cut applies to both the line-only
and line-on-continuum objects, regardless of whether this detected flux
is line or continuum. In doing so, we have a uniform cut in 
line flux across both samples.
We must ensure that these flux cuts define a region of near-100\%
detection probability. This time, the detectability is measured for
a sample of artificial objects with both line {\sl and continuum}.
A grid of objects with line and continuum values spanning the ranges in
Fig.~\ref{figlimits6}~({\em right}) was constructed using the same
worst-case noise values for Fig.~\ref{figlimits5}. The detection probability
as a function of these two parameters was obtained by measuring the
success rate of the software in finding such objects. The contours
in Fig.~\ref{figlimits6}~({\em right}) show levels of detection probability
in this space. The artificial objects were assumed to have the sky noise
associated with the largest available aperture for the field. 
As with the real object catalogue, they were only included if the
resulting line had a $\sigma$-deviation of 3 or greater.
The boundary between 2 and $3\sigma$-deviation
detections is represented in Fig.~\ref{figlimits6}~({\em right}) by
the dashed line. This is the locus of points with 3$\sigma$ line
fluxes in the largest aperture. Had we defined emission-line objects
at a lower level than 3$\sigma$, we would have populated the region
above this line with candidates as well.
However, in such cases it is difficult to be satisfied that
deviations less than $3\sigma$ are significant compared to the
natural scatter of points about the mean continuum level.

Our third and final cut is that representing the faintest limit to
which we can detect line flux relative to continuum. This is
an observed equivalent width (EW) cut, and the levels we
have chosen are shown by the labeled solid diagonal
lines in Fig.~\ref{figlimits6}~({\em right}). 
The definition for
equivalent width from narrowband measurements, is
\begin{equation}
{\rm EW} = \frac{F_{\rm line} \cdot \delta\lambda_{\rm e}} {F_{\rm cont}} .
\label{EWdefn}
\end{equation}
Here, $F_{\rm line}$ and $F_{\rm cont}$ are the respective line and
continuum fluxes and $\delta\lambda_{\rm e}$ is the effective width of the 
passband. The effective width, $\delta \lambda_{\rm e}$, is that width
of the rectangular bandpass with equal area to the true passband,
$\delta \lambda$, and with 100\%\ peak transmission. They are related
through $\delta \lambda_{\rm e} = \pi \cdot \delta \lambda / 2$. 
The {\sl rest-frame equivalent width} (REW) is that measured in the 
reference frame of the galaxy. It is related to the observed 
equivalent width by ${\rm REW} = {\rm EW}/(1+z)$, where $z$ is the 
galaxy redshift.

We make equivalent width cuts at a 
level ensuring that we are complete both in terms of detection probability
and 3$\sigma$ line detections. We know that we could have detected an
object anywhere within the boundaries of line and continuum flux, and
observed equivalent width. Table~\ref{t:selectlimits} lists the 
equivalent width cuts applied in each case. Also listed are the
rest-frame equivalent widths for \Ha\ at its redshift in these passbands.
We determine our overall sample completeness by comparing the
numbers of objects actually observed with those expected
over the same selection space. For each object we determined a
{\sl completeness weighting factor}, being the reciprocal of the
detection probability in the region of that object.
Most objects have weighting factors of 1.0 since they
lie in the region of 100\%\ detection probability for their respective
field. In summing the
completeness weights, each object is represented by the numbers we
{\sl should} have detected, thereby giving the total number of objects
we could have expected for the homogeneous sub-sample.
Figure~\ref{figcomplete2} shows the cumulative completeness of the homogeneous
sub-samples as a function of flux. Note that since completeness
is a function of surface brightness, its decline 
is not due to flux alone. However, 
Fig.~\ref{figcomplete2} demonstrates that all
samples are $\sim 93$\% complete or greater, at the limits we have set.
Table~\ref{t:homogeneous} gives the breakdown of types for
our homogeneous sub-sample, and the total numbers of observed and
expected objects.

\subsection{Flux Selection of an \Ha\ Sub-Sample\label{s:haSamp}}

Defining homogeneous emission-line samples
is insufficient on its own if we wish to characterize the
star-formation history of the universe. We expect our 
line-flux-selected samples to contain a combination of galaxies
seen through \Ha, \OII\ and \OIII\ at a
variety of redshifts matching the observed wavelength.
However, the conclusions of Sect.~\ref{s:volLimited} 
(in particular, Fig.~\ref{figcontamEVOL}) suggest that at the brightest 
fluxes, our samples will be dominated by \Ha\ emission-line galaxies.
The reason for this is as follows. Such fluxes translate
into luminosities brighter than $L^*$ in the emission-line
luminosity functions for the background galaxies.
As a consequence, \Ha\ galaxies dominate, being the only
set from which the members are significantly sampled beyond $L^*$. 
As can be seen in all cases of Fig.~\ref{figcontamEVOL}, the transition
from an \Ha-dominated sample to that inclusive of other lines
is quite sharp. This is due to the exponential decline in galaxy numbers
for a Schechter luminosity function for luminosities $L > L^*$.

The homogeneous sub-samples in the 814/33 and 909/40 filters
have fluxes greater than 0.71 and $0.57 \times 10^{-16}$~\lineunits\
respectively (Table~\ref{t:selectlimits}). 
By Fig.~\ref{figcontamEVOL}, we would
expect these samples to consist almost entirely of \Ha\
emission ($> 90$\%).
We define two \Ha\ samples for each of the 
707/26, 814/33 and 909/40 sets: one derived from the full sample
of emission-line objects and the other from the homogeneous
sub-sample. They are selected by taking only those galaxies
above the flux limit ensuring complete \Ha\ selection,
as determined by Fig.~\ref{figcontamEVOL}.
Table~\ref{t:halphaSamp} summarizes the flux cuts made and the
sizes of the resulting \Ha\ samples. The original
homogeneous sub-samples in the 814/33 and 909/40 filters
remain intact since the original flux limits
(0.71 and $0.57 \times 10^{-16}$~\lineunits\ respectively)
restrict these samples to \Ha\ galaxies almost entirely ($> 90$\%).
However, the homogeneous sub-sample in the 707/26 filter
has a wider mix of \Ha, \OII\ and \OIII\ galaxies
and so we take our \Ha\ sample as those with line fluxes exceeding
$1.50 \times 10^{-16}$~\lineunits. At this level the \Ha\
galaxies constitute more than 80\% of the total sample.
Table~\ref{t:halphaSamp} also lists the limiting line luminosities
for \Ha\ flux at the selection limit, at the mean \Ha\
redshift in each interval.

Although this dominance of \Ha\ at higher flux is fortuitous
for the present analysis, future work is necessary to
solve redshift ambiguity, in the form of follow-up spectroscopy. Not only
does this see the TTF emission-line identified, but the nature of the galaxy
emission and its reddening can be revealed through
ratios with other emission-lines.
Although such follow-up is beyond the scope of this paper,
it is the obvious direction for subsequent work on the survey.

An alternative solution to redshift ambiguity was explored
in the form of broadband imaging. 
As the major optical emission-lines are well-separated in
wavelength, the corresponding galaxy redshifts separate
the broadband colors unambiguously in a color-color plane.
This differs from conventional photometric redshift techniques
(see discussion by \cite{koo99}~1999) 
in that a {\sl precise} starting point for the redshifts
is given through the limited possibilities offered by \Ha\, \OII,
\OIII\ or \Hb. The disadvantage of this method is that it requires
{\sl all} emission-line galaxies (that make it into the final sample)
to have sufficient continuum for a broadband detection.
As the galaxies from our survey with sufficient continuum 
are few ($\sim 5$\%), they do not constitute a sample that both
meets this criterion and offers sufficient numbers to be useful.
Observations representing many hours on a 4~m-class telescope or 
greater are required, in which case, spectroscopic follow-up becomes an
increasingly feasible alternative.
We obtained broadband colors for 12 of the 15
{\sl TTF Field Galaxy Survey} fields. The regions were imaged in $BVRI$
on the 1.0~m telescope at Siding Spring Observatory, Australia.
This yielded usable colors for the subset of our galaxies
($\sim 13$\%) with sufficient continuum ($B \simlt 21$). 
Such colors can then be used to solve the galaxy redshift ambiguity.
More details on the application and results of this approach are presented
by \cite{jones99}~(1999). However, because the brightest galaxies were
too few to be useful, this method could not be applied as a redshift 
discriminant to the {\sl TTF Field Galaxy Survey} as a whole.


\section{SAMPLE CHARACTERISTICS\label{s:character}}

\subsection{Autofib Redshift Survey Comparison \label{s:autofib}}

We tested the TTF detection of emission-line galaxies against
a conventional broadband-selected redshift sample. The Autofib
redshift survey (\cite{ellis96}~1996) is a composite sample of 
1700 galaxies obtained through the combination of the 
DARS (\cite{peterson85}~1985), Autofib (\cite{ellis96}~1996),
BES (\cite{broadhurst88}~1988), LDSS-1 (\cite{colless90}~1990, 1993)
and LDSS-2 (\cite{glazebrook95}~1995) samples.
Galaxies have been selected from the magnitude range
$ 11.5 < b_J < 24.0$ and the presence of emission is judged through
the \OII\ line.\footnote{The $b_J$ passband is defined by the 
combined wavelength sensitivities of the Kodak IIIa-J emulsion and a
GG385 glass filter.
} 
The different samples probe different magnitude
ranges and areas on the sky.
There are 827 members of the Autofib redshift survey that
can be classified as emission-line galaxies, defined as such by \OII\
rest-frame equivalent widths greater than 10~\AA.
Of these, there are 18 within one of our 9 overlap fields at redshifts
placing \Ha\ within the relevant scan window (707/26, 814/33 or 909/40). 
We examined the detection success of TTF on these galaxies as a 
function of both $b_J$ magnitude and \OII\ rest-frame equivalent 
width.

Table~\ref{t:figautofib} summarizes the results. In addition to the 
TTF detection rate, we show how many of the TTF object detections occurred
as line-on-continuum and how many were line alone.
Table~\ref{t:figautofib} shows the trend of
fainter galaxies ($b_J \simgt 21$) becoming \Ha\  line-only
detections in increasing numbers. This we expect since the narrow
TTF passband is optimized for the detection of faint line flux
and not faint continuum. There is also the suggestion of a declining
detection rate with fainter limiting apparent
magnitude, although it is difficult to tell with such small
numbers of galaxies. {\sl All} of the undetected galaxies (except one)
lie at $b_J \sim 21.0$. The one fainter $b_J \sim 22.0$ galaxy
which was detected, was detected as a line-only detection.
We conclude from this that $b_J = 21.0$ is around the limit at which we 
can detect galaxies in both their line and continuum. Beyond this,
line detections dominate.
When ranked according to \OII\ rest-frame equivalent width, little is 
seen in the way of TTF detection trends. Again, the small size of the
overlap sample is a limitation, particularly because the majority of 
objects have \OII\ emission with low to moderate rest-frame equivalent 
widths ($10 \simlt {\rm EW} \simlt 40$~\AA).

We have assumed here, however, that all galaxies
exhibiting \OII\ emission will have equally strong \Ha.
Although this is commonly true it is not always the case, and some
of the TTF non-detections in Table~\ref{t:figautofib} may have been
simply because there was little or no \Ha.
\cite{tresse99}~(1999) find a small but significant fraction of the
Stromlo-APM Survey (11\%) that show \OII\ but no \Ha.
They attribute this to strong stellar absorption and do not find
such galaxies to be more prevalent at a particular redshift. 
So an alternative possibility is that Table~\ref{t:figautofib} underestimates
the detection success of TTF in some cases. However, this 11\% is less
than the $\sim 25$\% of Autofib \OII\ galaxies that TTF does not detect.
This suggests that an additional factor is responsible, although 
it is difficult to be certain when such small sample sizes are involved.
\cite{tresse99} note that the fraction
of their broadband-selected sample
seen in emission ($\sim 60$\%) is the same, irrespective
of whether emission is defined through the presence
of \OII\ or \Ha. Remarkably, they find that the number of galaxies
with \OII\ and no \Ha\ negates the galaxy numbers with \Ha\ and no \OII.

Figure~\ref{figautofib} shows the relationship between
spectral quantities measured around \Ha\ by TTF and \OII\ by
the Autofib survey for the galaxies in common. In panels
($a$) and ($b$), we find little correlation between either the
\Ha\ and \OII\ emission-line fluxes or between the corresponding
equivalent width measures. We do not see the
EW(\Ha$+$\NII) versus EW(\OII) correlation of \cite{kennicutt92a}~(1992a)
or \cite{tresse99}~(1999), probably because this correlation is
weak with large scatter and our sample size is small.
The large scatter in Fig.~\ref{figautofib}($b$) may also be due to
a tendency of spectroscopic equivalent widths to be
overestimates at the limits of a broadband-selected sample, as
discussed in Sect.~\ref{s:sample}.
Either way, our results lend further weight to
\cite{kennicutt92a}~(1992a) who finds \OII\ flux and
equivalent width to be loosely connected
to the star-formation rate in a galaxy.
In Fig.~\ref{figautofib}($c$) we show the relationship between
the Autofib $b_J$ magnitudes and the TTF narrowband continuum 
measurements. There is reasonable agreement between the two
with the line of best-fit being
\begin{equation}
b_J = -2.5 \log F({\rm cont}) + 21.81 ,
\label{contFit}
\end{equation}
where $b_J$ is the photographic magnitude and $F({\rm cont})$ is the
narrowband continuum in the TTF bandpass ($10^{-16}$ \ergs2band).
In making the fit we fixed the slope to $-2.5$ and weighted the points
by the $1/\Delta F^2$ continuum photometry uncertainties.
The agreement in panel ($c$) indicates that the scatter in the
equivalent width measurements of panel ($b$) is due primarily to the
dispersion in line flux seen in panel ($a$).
Equation~(\ref{contFit})
implies that the {\sl continuum} detection limit of TTF is
about the same as the $b_J = 24.0$ limit of the broadband-selected survey.

There is little overlap between the
emission-line selected galaxies found in the
the broadband-selected Autofib redshift survey and the
{\sl TTF Field Galaxy Survey}. The vast majority of TTF-selected
galaxies have continua too faint for inclusion in the broadband
catalogue while the redshifts of the Autofib line-emitters are
too wide-ranging for the relatively narrow scan windows of TTF.
For the galaxies in common, TTF detects the brighter Autofib
galaxies ($b_J \simlt 21$) in both line and continuum while
fainter line-emitters are almost exclusively detected in the
line alone. While there is good agreement between the
redshifts and continuum fluxes measured by TTF and Autofib around
\Ha\ and \OII, there is considerable scatter between the
measurements of line flux and equivalent width.
However, much of this dispersion can be attributed to intrinsic
differences between the \Ha\ and \OII\ lines themselves,
as seen in the loose correlations between these lines by
\cite{kennicutt92a}~(1992a) and \cite{tresse99}~(1999).
Improved insight into the relative merits of TTF
over broadband-selected emission-line samples would be obtained
through a redshift survey with near-infrared spectra
overlapping with the TTF wavelength intervals directly,
so that the same emission-line could be observed by each technique.

\subsection{Emission-Line Distributions\label{s:sample}}

Previously (Sect.~\ref{s:strategy}) we discussed the need to examine
the emission-line distributions with wavelength for signs of clustering.
In the case of our \Ha\ sub-samples we can interpret the
wavelength distributions directly as distributions in redshift
space. Figure~\ref{figpeaks2} shows such distributions by field for 
galaxies in the entire \Ha\ sub-sample.
No clear clustering is evident, although there are suggestions of
peaks in the 14\_3A, 22\_3A and 13\_3B fields. The 22\_3A clustering
is irrelevant because this field has been excluded from the homogeneous
samples used in our final analysis. 
Certainly we find no strong evidence for clusters in the 10\_3C, 10\_3F,
10\_3H or 13\_3E fields which coincide (directionally at least) 
with the \cite{lidman96}~(1996) candidates.
Given the limited impact that clustering appears to have on our
field samples, we leave them intact as representative samples of 
the field population. 
Of greater interest are the fluctuations in the total numbers of \Ha\
galaxies between individual fields, particularly in the 707 set, and to
a lesser extent in 814. This is due to the high flux cut imposed upon the 
707 sample to ensure a sample of \Ha\ galaxies. At this level, the 
numbers of galaxies field-by-field are sufficiently few that
fluctuations of this order are inevitable. 
Such variations are also seen between fields of broadband-selected 
redshift surveys (cf.~redshift spikes in Fig.~3 of 
\cite{broadhurst88}~1988).

For Fig.~\ref{figpeaks2}, the peak wavelength of an emission-line 
was determined through two different methods depending
on the nature of the emission. In objects with no continuum, the
peak wavelength was that of the slice showing emission,
or the mean wavelength if two adjacent slices showed emission.
In objects with continuum, the peak wavelength of the most deviant
slice, or the mean wavelength of two adjacent deviant slices was used.
Mean wavelengths were computed by weighting the wavelengths
of individual slices by the line flux in each.
The mean flux ratio \Ha/(\Ha$+$\NII)~$= 0.69$ (\cite{tresmad98}~1998) 
was used to adjust \Ha\ fluxes for the contribution of \NII.
In all cases, wavelength corrections were applied to account
for the phase effect with object position relative
to the optical centre of the Fabry-Perot beam.

Table~\ref{t:emisstype} summarizes the occurrence of emission-lines
in one and two slices for the homogeneous sub-sample of candidates.
The cases of detection and non-detection of continuum are shown
separately. Single-slice emission-lines represent $\sim 50$\% 
of objects with continuum, and a much larger fraction ($\sim 98$\%)
of those objects without. We would expect a reasonable fraction
of the emission-lines to appear in two, rather than one slice,
given the spacing and width of the slices. This is true for the
line-on-continuum detections. The extreme ratio of one to two-slice
candidates for line-only case suggests that many of the
one-slice objects could be noise peaks. However, not all are,
as Table~\ref{t:figautofib} shows with the comparison to the
Autofib data. Only follow-up multi-object 
spectroscopy can resolve this difficult but crucial issue.

\subsection{Number Count and \Ha\ Luminosity Function Evolution\label{s:numberCounts}}

In this section we examine the evolution in the number count and
volume-averaged star-formation rates of the {\sl TTF Field Galaxy Survey}. 
From an observational perspective we would like some measure of the density 
of emission-line sources on the sky. This should comprise all 
sources with redshifts placing the emission-line within the observed 
wavelength interval, whatever that emission-line may be. 
Knowing this quantity is fundamental to optimizing source 
coverage and maximizing sample size for a volume-limited 
emission-line survey (Sect.~\ref{s:strategy}).
Multi-object spectroscopic follow-up to narrowband samples
also requires a knowledge of object densities when deciding between 
strategies of spectral coverage and slit placement. 
From the perspective of galaxy evolution, projected number counts 
potentially provide a strong constraint on evolutionary models, 
since they comprise galaxies from more than one redshift 
bin simultaneously. Thus a chosen model must reproduce both the combined 
numbers observed at each wavelength and the relative mix of \Ha,
\OII, \OIII\ and other galaxies, all subject to the same evolution.
As such, narrowband number counts should provide a stringent test
for competing evolutionary models of the star-forming population.

Figure~\ref{fignumbersEVOL2} shows the cumulative narrowband number counts
from the complete homogeneous sub-sample of the {\sl full} emission-line 
set (including \Ha). Sample cuts at 4 and 5$\sigma$ demonstrate
how the weakest detections influence the counts obtained. 
The sigma values in this case are the line strengths expressed as a
statistical deviation above the background, be it sky or
continuum noise. 
Corrections for incompleteness have been applied
to all of the counts shown in Fig.~\ref{fignumbersEVOL2}. Counts are shown
down to the selection limits of the homogeneous sub-sample
(Table~\ref{t:selectlimits}) and the flux cuts for the \Ha\ 
sub-samples are indicated by the dashed lines.
Counts are expressed in terms of the projected density of emission-line
objects on the sky per bin per 100~\AA\ wavelength scan coverage.
In this way, we have a useful measure of the object counts obtained
in a flux-limited tunable filter survey, 
in terms of quantities directly observed.
Note that the 100~\AA\ is {\sl not} the scan interval (as given by 
column 2 of Table~\ref{t:scans}), but rather the product of the
effective bandpass and the number of slices per scan.
Also observe that 100~\AA\ represents a different redshift coverage
$\Delta z$ at the rest wavelength of each major optical emission-line.
However, it is fixed for that line, irrespective
of the redshift at which it is observed.

The {\em solid curves} in Fig.~\ref{fignumbersEVOL2}
show the estimated numbers we should have expected based on
evolving \Ha\ luminosity function calculation, of the type
detailed in Sect.~\ref{s:volLimited}. 
Thus, the
number evolution assumes our earlier model constrained by the
\cite{gallego95}~(1995) and \cite{tresmad98}~(1998) \Ha\ luminosity
functions, namely
\begin{equation}
\begin{array}{c}
\log \phi^* (z) = -3.2 + 4.68 \,\log_{10}(1+z), \\
\log L^* (z) = 42.15 - 0.25 \,\log_{10}(1+z), \\
\alpha(z) = -1.3 \, - 0.25 z . \\
\end{array}
\label{model1}
\end{equation}
This is identical to the general form presented in Eqn.~(\ref{phiLwithZ}).
These surveys were chosen due to both their spread in redshift and
use of galaxy spectra. The contribution from other major emission-lines
is calculated by scaling the \Ha\ luminosity function by the
appropriate line-ratio ({\em c.f.}~Fig.~\ref{figkenevol}) 
at the required redshift.
Note that this is {\sl not} a model of our observed counts, but
a comparison of those we could have expected based on the luminosity
functions of the earlier surveys.

Figure~\ref{fignumbersEVOL2} shows that the model expected counts
({\em solid line}) are a fair match to the $\ge 5\sigma$  
distributions in the 707/26 and 814/33 intervals. In the 
909/40 interval they are a much closer match to the $\ge 4\sigma$
sample. There are two potential contributions to this difference.
The most likely is that we are simply less-sensitive
to the same type of galaxies at higher redshift, rendering what
we detect as $5\sigma$ in 707/26 and 814/33, as $4\sigma$ in
909/40. A second effect could be that the model constraints at 
$z \sim 0$ (\cite{gallego95}~1995)
and $z \sim 0.2$ (\cite{tresmad98}~1998) are insufficient for the
extrapolation to $z \sim 0.4$ where \Ha\ is seen in the 909/40 interval.
However, there are presently no \Ha-selected galaxy surveys that
encompass $z \sim 0.4$.

Observe also in Fig.~\ref{fignumbersEVOL2} the significant excess 
($\sim 2$ to 3 times) in the faintest narrowband number counts 
over the comparison with previous surveys ({\em solid line}).
While we have not proven the validity of all detections less
than $5 \sigma$, our visual confirmation of all line-on-continuum
candidates (13\%) represents a greater fraction of the total counts
than does the {\em solid line}, at the faintest limits where the 
difference is largest. This suggests we have an excess in
emission-line galaxies over the numbers we would otherwise estimate
from these two previous \Ha\ surveys. The corollary of this is
that the population of faint star-forming galaxies is more numerous
than these surveys have found, consistent with the preliminary
findings of Gronwall~\etal~(priv.~comm.) for a new wide-field \Ha-survey
at $z \sim 0.08$. Such a scenario is also consistent 
with our conclusions from the comparisons
between the TTF and Autofib samples in Sect.~\ref{s:autofib}.

Before interpreting such higher star-formation densities as indicative
of the field, we should consider possible sources of contamination.
Both active galactic nuclei (AGN) and quasi-stellar objects (QSOs)
are potential emission-line sources at non-zero redshift. 
Like \cite{tresmad98}~(1998), we have not removed
AGN as their influence on the general emission-line population has
previously been shown to be small ({\em i.e.}~$\le 10$\%; 
\cite{sarajedini96}~1996; \cite{tresse96}~1996).
Likewise, the occurrence of QSOs to be rare against
the general star-forming population: $\le 2$~\persqdeg\ 
from within a typical high-$z$ TTF interval of
$\Delta z = 0.1$, using the number-redshift distribution
of \cite{boyle88}~(1988).
This is consistent with the wide-field objective prism survey 
of \cite{smith76}~(1976) for emission-line objects
down to $b_J \simlt 18$, with 96\% as
galaxies and 2\% as possible QSOs. 
The most likely sources of Galactic contamination are late-type stars. 
\cite{boroson93} (1993) in their 5-filter intermediate-band imaging survey
using $80$~\AA\ bandpasses, attributed $\sim 10$ to 20\% of
their {\sl rejected} emission-line candidature to late-type stars.
Such stars have strong molecular TiO absorption bands
with widths of typically $\sim 500$~\AA.
Presumably these spectral variations were difficult to distinguish
from genuine line-emission in the broader $80$~\AA\ bandpasses used by 
these authors. However, our narrower bandpasses ($\sim 20$ to 30~\AA), the 
greater number of them (10), and smaller scan interval ($200$ to 300~\AA),
argues that we should not be susceptible to this problem in the 
same way.

Given the unlikelihood of celestial contamination,
our data suggest greater numbers of star-forming galaxies than 
has previously been attributed to the field population.
A consequence of this would be higher mean star-formation densities,
although without confirmation of our faintest detections through additional
observations, we can not quantify the increase.
Specifically, follow-up spectroscopy is required to 
provide independent confirmation of the faintest emission-line
detections, in addition to confirmation of the line as \Ha.
\cite{baker01}~(2001) have obtained follow-up spectroscopy to
a sample of TTF-selected emission-line galaxies in the field
of QSO MRC~B0450$-$221 at $z = 0.9$. Of the 5 TTF candidates targetted,
3 were positively identified as emission-line sources, and 1 as a possible
emission-line detection. This important result
demonstrates the capacity of TTF for detecting faint emission-line sources
at high redshift.

A more immediate approach to evolutionary trends is to 
compare \Ha\ luminosity functions directly. Figure~\ref{figLFdn}
shows the luminosity functions of our homogeneous
\Ha\ sub-samples in each interval for the $\ge 4\sigma$ 
({\em filled circles}) sample. The \Ha\ luminosity functions of 
\cite{gallego95}~(1995; {\em open triangles}) and 
\cite{tresmad98}~(1998; {\em open circles}) are also shown in the
relevant redshift interval. All luminosity functions are shown
uncorrected for extinction. Poissonian errors derived from the square-root 
of the galaxy numbers in each bin are shown. For the TTF data,
\Ha\ fluxes were corrected for the effect of
\NII\ using the mean flux ratio \Ha/(\Ha$+$\NII)~$= 0.69$, obtained by
\cite{tresmad98}~(1998) for their sample at $z \sim 0.2$.
Star-formation rates were derived using the same calibration as
adopted by \cite{glazebrook99}~(1999), namely
\begin{equation}
\frac{{\rm SFR}} {( M_{\odot}\, {\rm yr}^{-1} )} = 7.4 \times 10^{-42}
\frac {L_{\rm H\alpha}} {( {\rm erg} \, {\rm s}^{-1})}
\label{SFRha2}
\end{equation}
from the \cite{bruzchar96}~(1996) models with solar metallicity and
a Salpeter IMF. No extinction corrections have been applied to
the luminosities, although corrections have been 
made for the effect of the Airy profile on sampling an ensemble
of galaxy line fluxes (\cite{jones99}~1999). 
We save the derivation of a luminosity function fit for when 
we have spectroscopic confirmation of the \Ha\ sample in hand.

Figure~\ref{figLFdn}($a$) demonstrates the excess in our $z = 0.08$ counts
over the faint-end of the \cite{gallego95}~(1995) distribution
by almost an order of magnitude. Much closer agreement exists between
our values and those of a preliminary \Ha\ luminosity function from
the KPNO International Spectroscopic Survey (KISS; Gronwall~2000, 
priv.~comm). Taken together, these results 
suggest that the \cite{gallego95} survey
underestimates the faint-end of the local \Ha\ luminosity function,
and hence, the star-formation content of the local universe.
For the $z=0.24$ interval (Fig.~\ref{figLFdn}$b$) we find broad
agreement with the \Ha\ luminosity function derived by
\cite{tresmad98}~(1998). Due to the volume-limited nature of the survey
we have insufficient numbers to define the bright-end ``knee''
of the distribution. Our measurements at $z=0.40$
(Fig.~\ref{figLFdn}$c$) are the first to use \Ha\ in the
$\sim 5$~Gyr of lookback time separating the samples of
\cite{tresmad98}~(1998) at $z \sim 0.2$ and
\cite{glazebrook99}~(1999) at $z \sim 0.9$.

Our preliminary \Ha\ luminosity functions argue for an upwards
revision of the star-formation content for the local universe.
This in turn implies a less-dramatic decline in the cosmic star-formation
history than has been shown to date
(\cite{tresmad98}~1998; \cite{glazebrook99}~1999; \cite{yan99}~1999). 
However, follow-up spectroscopy is needed to confirm the nature
of the \Ha\ detections and to correct individual \Ha\ fluxes for
the effects of extinction and \NII.
Indeed, if many of our sub-$4\sigma$-detections are verified, then
the measured excess would be greater.


\section{CONCLUSIONS\label{s:conclremarks}}

We now summarize our preliminary findings from a census of the
{\sl TTF Field Galaxy Survey} for star-forming galaxies.

\begin{enumerate}

\item Using the {\sc Taurus} Tunable Filter (TTF), we have
obtained a narrowband-selected sample of emission-line galaxies
in the field. The survey is composed of 15 scans of 10 narrowband
slices covering 0.27~\sqdeg\ on the sky, with 7 of the scans in a 
707/26~nm wavelength interval, 5 in 814/33~nm and 3 in
909/40~nm. There are 
696 candidates (at the $3\sigma$ level or greater) in a homogeneous 
sub-sample for which we
can correct for incompleteness (96.4\%). Although some detections
(13) are confirmed by the Autofib Galaxy Redshift Survey,
the majority exist without separate confirmation.
This is required for both checking the validity of detections
and correctly identifying the emission-line concerned.
The homogeneous sub-sample
excludes two of the worst fields, one in each of the 707 and 909 
intervals. Through a synthesis of \Ha\
luminosity functions and median galaxy emission-line ratios,
we estimate the relative occurrence of different emission-lines
in our sample. From this we expect our samples to be 
dominated by \Ha\ galaxies in the sub-samples with line fluxes
exceeding 1.50, 0.71 and $0.57 \times 10^{-16}$~\lineunits\ 
for the 707, 814 and 909 intervals respectively. 
In these flux regimes, we have 
660 candidates from the homogeneous sub-sample
that we expect to consist mostly of \Ha\ galaxies,
correctable for the small amount of incompleteness (96.5\%)
inherent.

\item We find little evidence for strong clustering in either
the full emission-line sample or the \Ha\ subset. Those of our
fields overlapping with the \cite{lidman96}~(1996) cluster 
candidates did not yield any clear peaks in redshift space.
This allows us to rule out clusters in our field sample
as seen in emission-lines. It does not, of course, tell us
anything about possible clusters outside of our scan intervals
but in the same direction on the sky.

\item We compared the detection ability of TTF in finding emission-line
galaxies to the broadband-selected Autofib galaxy redshift survey
(\cite{ellis96}~1996).
Our comparison was made between the \OII\ galaxies found in Autofib
and the \Ha\ galaxies of TTF. There is little overlap between the
emission-line samples of the two surveys. For the sky regions
that overlap, most of the TTF-selected galaxies
are not seen in the Autofib catalogue because they have continua
beyond the selection limit of this catalogue. Conversely,
TTF misses many of the Autofib emission-line galaxies because
they are beyond the comparatively narrow volumes sampled by TTF.
For the galaxies in common, TTF detects $b_J \simlt 21$ galaxies in both line
and continuum while fainter galaxies are detected in the line alone.
TTF recovered 13 of an expected 18 Autofib emission-line galaxies,
although those that were missed may have been due to weaker \Ha\
emission than \OII. 
The two surveys agree in their continuum and redshift measurements
although show scatter in their line flux and equivalent width measures.
However, much of this scatter may be due to intrinsic scatter 
in the correlation of \Ha\ and \OII\ fluxes. Better insight into
the detection rates of the two methods would be gained through
a direct comparison concentrating on the same emission-line at
wavelengths common to both.

\item The projected density of emission-line galaxies on the
sky was compared to what we would have expected from existing
measurements of the \Ha\ luminosity function. At the faintest
limits we find a significant excess in these narrowband 
number counts, up to 3 times the expected counts at the faintest
flux limits of our $\ge 4\sigma$ sample. This excess is also
seen when our preliminary \Ha\ luminosity functions are 
compared directly to those of other surveys, in each redshift
interval individually. Although the influence of false detections
for the line-only candidates is yet to be quantified, 
our visual confirmation of the line-on-continuum candidates (13\%)
is consistent with an excess of emission-line candidates.
Such an excess is most likely due to our narrowband selection
of the sample, which is without precedent for field emission-line
galaxies at these redshifts. The greater numbers imply
higher star-formation densities than have previously been
attributed to these epochs. However, follow-up measurement of the
detections is required to quantify this increase.

\end{enumerate}

Multi-object spectroscopic follow-up is necessary for 
future work on this sample. Such spectra would cover a wider
wavelength range than TTF, and as such, would allow us to resolve 
many of the issues for which we have presently made assumptions.
Spectroscopic information would first allow independent confirmation
of the line detections and resolve the redshift ambiguity.
Furthermore, with spectra we can determine the
amount of reddening in individual galaxies through the \Ha/\Hb\
ratio and correct each for such. If the spectra are of sufficiently 
high spectral resolution, the \Ha\ and \NII\ lines can be separated, 
thereby avoiding contamination of the former by the latter. 
Furthermore, the \Ha\ line fluxes 
could be measured directly from the line which would remove the 
need for the bandpass correction. Finally, through an assessment
of ratios such as \NII/\Ha\ and \OIII/\Hb\ we could identify and reject
AGN from our sample of star-forming galaxies
(e.g.~\cite{veilleux87}~1987). Many existing
\Ha\ surveys (e.g.~\cite{tresmad98}~1998; \cite{yan99}~1999)
include these emission-line sources, although they occur in
small numbers relative to the general star-forming population.

In short, the {\sl TTF Field Galaxy Survey} is a survey for star-forming
galaxies is largely unprecedented in its systematic narrowband seach for field
galaxies. Our data suggest galaxy counts and mean star-formation rates higher 
than those measured by previous \Ha\ and UV surveys.
An intriguing possibility for the higher values we observe is
a population of faint star-forming galaxies, with continua sufficiently
faint to place them beyond the range of conventional redshift surveys.
However, follow-up measurements utilizing multi-object spectroscopy
are required for this to be confirmed.

\acknowledgments
It is a pleasure to thank Matthew Colless for many helpful discussions
regarding this work. We are also grateful to the referee Caryl Gronwall
for her reading of the original manuscript and the many improvements
that resulted from her suggestions. The comments of the referees 
who examined this work during its presentation for a Doctoral Thesis are also
gratefully appreciated. We acknowledge the Australian Time Allocation
Committee for their support of the TTF observations at the AAT upon 
which this work is based. This research has made use of the NASA/IPAC
Extragalactic Database (NED), which is operated by the Jet Propulsion
Laboratory, Caltech, under contract with NASA.
DHJ acknowledges the support of an Australian Postgraduate 
Research Award during the completion of this work.



\clearpage
\begin{deluxetable}{lcccc}
\tablewidth{0pc}
\tablecaption{Scan Parameters\label{t:stepping}}

\tablehead{
\colhead{Filter $\lambda/\Delta \lambda$}	& 
\colhead{Range of Central} & 
\colhead{Passband}	& 
\colhead{Effective} &
\colhead{Step size} \\
\colhead{(nm)} & 
\colhead{Wavelengths (\AA)} & 
\colhead{FWHM (\AA)}	& 
\colhead{Passband (\AA)} &
\colhead{(\AA)} \\
}
\startdata
707/26 & 6986--7183 & 12.9 & 20.3 & 21.9 \nl
814/33 & 8026--8263 & 16.4 & 25.8 & 26.3 \nl
909/40 & 8945--9245 & 22.3 & 35.1 & 33.4 \nl
\enddata
\end{deluxetable}

\clearpage

\begin{deluxetable}{lcccccccc}
\scriptsize
\tablewidth{0pc}
\tablecaption{Narrowband Scan Coverage\tablenotemark{} \label{t:scans}}

\tablehead{
\colhead{Filter $\lambda/\Delta \lambda$}	& 
\colhead{Redshift Range} & 
\colhead{$\bar{z}$}	& 
\colhead{Proper Distance} &
\colhead{$\delta v$ \tablenotemark{a}} &
\colhead{No.\ of} &
\colhead{Total Volume} &
\colhead{$\log(L_{\rm line})$ \tablenotemark{b}} &
\colhead{SFR \tablenotemark{b} \tablenotemark{c}} \\
\colhead{(nm)} & 
\colhead{($z_1 \leq z \leq z_2$)} & 
\colhead{ }	& 
\colhead{(\Mpc3)} &
\colhead{(\kmPerSec)} &
\colhead{Fields} &
\colhead{(\Mpc3)} &
\colhead{(\ergsPersec)} &
\colhead{(\Msunyr)} \\
}
\startdata
\Ha\ line: &	&	&	&   &     &      &      &     \nl
707/26 & $ 0.062 \leq z \leq 0.093 $  & 0.077  & 438  & 8610  & 7  & 1170  & 39.43  & 0.02  \nl
814/33 & $ 0.221 \leq z \leq 0.260 $  & 0.24  & 1230  & 9510  & 5  & 6660  & 40.44  & 0.22  \nl
909/40 & $ 0.359 \leq z \leq 0.411 $  & 0.39  & 1800  & 11200 & 3  & 9850  & 40.87  & 0.59  \nl
	&	&	&	&     &   &      &      &     \nl
\OII\ line:\tablenotemark{d} &	& 	&	&     &   &      &      &     \nl
707/26 & $ 0.870 \leq z \leq 0.924 $  & 0.90  & 3290  & 8610  & 7  & 49500  & 41.67  & 6.57 \nl
814/33 & $ 1.149 \leq z \leq 1.219 $  & 1.18  & 3880  & 9510  & 5  & 50300  & 41.93  & 12.14 \nl
909/40 & $ 1.393 \leq z \leq 1.484 $  & 1.44  & 4320  & 11200  & 3  & 42500  & 42.12  & 18.73 \nl
	&	&	&	&     &   &      &      &     \nl
\enddata
\tablenotetext{}{Assuming $H_0 = 50$ \kmsMpc and $q_0 = 0.5$. Circular TTF field of 9\am\ diameter.}
\tablenotetext{a}{velocity coverage per field in the rest frame of the emission-line sources, $\delta v = c(z_2-z_1)/(1+\bar{z})$.}
\tablenotetext{b}{\Ha\ or \OII\ log line luminosity at a detected flux of 
$1 \times 10^{-16}$ \lineunits.}
\tablenotetext{c}{star-formation rate using SFR (\Msunyr) $= L_{{\rm H}\alpha}/10^{41.10}
= L_{\rm [O~II]}/10^{40.85}$, with luminosities expressed in (\ergsPersec),
excluding internal galactic extinction.}
\tablenotetext{d}{not as prevalent as \Ha\ at the flux limits of the survey.}
\end{deluxetable}

\clearpage
\begin{deluxetable}{ccccccc}
\scriptsize
\tablewidth{0pc}
\tablecaption{{\sl TTF Field Galaxy Survey}: Log of Observations \label{t:observations}}

\tablehead{
\colhead{Field}	& 
\colhead{R.~A.\ (J2000)} & 
\colhead{Dec.\ (J2000)} & 
\colhead{Number}	& 
\colhead{Exposure Time} & 
\colhead{Observation} &  
\colhead{Autofib}\\
\colhead{ }	& 
\colhead{ (~$^{\rm h}$~~$^{\rm m}$~~~$^{\rm s}$~)} & 
\colhead{ (~\d~~\am~~\as~)} & 
\colhead{of Slices}	& 
\colhead{(s~slice$^{-1}$)} &  
\colhead{Date} &  
\colhead{Galaxies\tablenotemark{a}}\\
}
\startdata
707/26 scans:	&    &   &			&	&	&   \nl
00\_3A &  00~54~48.0 & $-$30~25~00 &  9   &   600 &  1997 Oct 08 &   2  \nl
10\_3F &  10~45~18.0 & $-$00~05~00 &  10  &   600 &  1997 May 01 &   -- \nl
10\_3D &  10~45~42.0 & $-$00~15~00 &  10  &   620 &  1998 Apr 04 &   2  \nl
12\_3A &  12~44~18.0 & $-$00~07~00 &  10  &   600 &  1997 May 01 &   -- \nl
13\_3E &  13~44~12.0 & $-$00~01~00 &  10  &   600 &  1998 Apr 04 &   2  \nl
14\_3A &  14~45~30.0 & $+$00~05~00 &  10  &   400 &  1997 Apr 30 &   -- \nl
       &	     &	           &  10  & $+$200 &  1997 May 01 &   -- \nl
22\_3A &  22~05~10.9 & $-$18~35~00 &  10  &   400 &  1997 Apr 30 &   -- \nl
       &	     &	           &  10  & $+$200 &  1997 May 01 &   -- \nl
	&	&   &			&	&		&   \nl
814/33 scans:	&    &   &			&	&	&   \nl
01\_3A &  01~05~48.0 & $-$30~00~00 &   9  &   600 &  1997 Oct 08 & 	 1  \nl
13\_3C &  13~43~27.0 & $-$00~20~00 &  10  &   620 &  1997 Apr 30  & 	 2  \nl
13\_3B &  13~44~04.8 & $-$00~22~00 &  10  &   600 &  1998 Apr 05 &   2  \nl
13\_3E &  13~44~12.0 & $-$00~01~00 &  10  &   600 &  1998 Apr 04 &   2  \nl
14\_3A &  14~45~30.0 & $+$00~05~00 &  10  &   600 &  1997 May 01 &   -- \nl
	&	&   &			&	&		&   \nl
909/40 scans:	&    &   &			&	&	&   \nl
10\_3C &  10~44~30.0 & $-$00~07~00 &   9  &   1080 & 1998 Apr 05 &   -- \nl
10\_3H &  10~46~24.0 & $-$00~24~00 &  10  &   1080 & 1998 Apr 03 &   2  \nl
13\_3D &  13~44~42.0 & $-$00~11~30 &  10  &   1080 & 1998 Apr 03 &   3  \nl
\enddata
\tablenotetext{a}{Emission-line galaxies from the Autofib survey
(\cite{ellis96} 1996) with (1) redshifts placing \Ha\ within the scan interval,
and, (2) \OII\ rest-frame equivalent widths $\ge 10$ \AA.}
\end{deluxetable}

\clearpage
\begin{deluxetable}{ll}
\tablewidth{0pc}
\tablecaption{\Ha\ Luminosity Function Evolution\label{t:calculations}}

\tablehead{
\colhead{Luminosity Function}	& 
\colhead{Evolutionary} \\
\colhead{at $z=0$}	& 
\colhead{Indices} \\
}
\startdata
$\phi_0^* = 0.63 \times 10^{-3}$ \MpcPer3  &   $\gamma_\phi = 4.68$  \nl
$L_0^* = 10^{42.15}$ \ergsPersec  &   $\gamma_L = -0.25$  \nl
$\alpha_0 = -1.3$ &   $\gamma_\alpha = -0.25$  \nl
\enddata
\end{deluxetable}

\clearpage
\begin{deluxetable}{ccccc}
\tablewidth{0pc}
\tablecaption{Selection Limits of Homogeneous Sub-Sample\label{t:selectlimits}}

\tablehead{
\colhead{Filter} &
\colhead{Flux}	& 
\colhead{Central Surface} &
\colhead{EW} & 
\colhead{REW in \Ha} \\ 
\colhead{$\lambda/\Delta\lambda$} &
\colhead{(\lineunits)}	& 
\colhead{Brightness\tablenotemark{a}} &
\colhead{(\AA)} & 
\colhead{(\AA)} \\
\colhead{(nm)}	& 
\colhead{}	& 
\colhead{(\lineSB)} &
\colhead{} & 
\colhead{} \\ 
}
\startdata
707/26  & $1.2 \times 10^{-16}$   &    $0.031 \times 10^{-16}$  &    2.2   &    2.0   \nl
814/33  & $0.71 \times 10^{-16}$  &    $0.031 \times 10^{-16}$  &    4.0  &    3.2   \nl
909/40  & $0.57 \times 10^{-16}$   &    $0.031 \times 10^{-16}$ &    4.0   &    3.3  \nl
\enddata
\tablenotetext{a}{Assuming an exponential disk profile}
\end{deluxetable}

\clearpage
\begin{deluxetable}{lccccc}
\tablewidth{0pc}
\tablecaption{Homogeneous Sub-Sample\label{t:homogeneous}}

\tablehead{
\colhead{Filter $\lambda/\Delta \lambda$}	& 
\colhead{Line-only} & 
\colhead{Line-on-Continuum}	& 
\colhead{Total No.} &
\colhead{Total No.} &
\colhead{Completeness\tablenotemark{a}} \\
\colhead{(nm)} & 
\colhead{Objects} & 
\colhead{Objects}	& 
\colhead{Observed} &
\colhead{Expected} &
\colhead{(\%)} \\
}
\startdata
707/26	&  75	& 30		& 105	& 111.89		& 93.8 \nl
814/33	& 387	& 36		& 423	& 431.76		& 98.0 \nl
909/40	& 141	& 27		& 168	& 178.42		& 94.2 \nl
	& 	&		& 	& 		&  \nl
Overall: & 603	& 93		& 696	& 722.07	& 	96.4 \nl
\enddata
\tablenotetext{a}{over the entire homogeneous sub-sample in each filter}
\end{deluxetable}

\clearpage
\begin{deluxetable}{ccccc}
\tablewidth{0pc}
\tablecaption{\Ha\ Sub-Samples From Full and Homogeneous Sets\label{t:halphaSamp}}

\tablehead{
\colhead{Filter}	& 
\colhead{Limiting \Ha\ Flux} &
\colhead{$\log L({\rm H}\alpha)$\tablenotemark{a}} &  
\colhead{From Homog.} &
\colhead{No.\ Expected} \\
\colhead{(nm)} & 
\colhead{(\lineunits)}	& 
\colhead{(\ergsPersec)}	& 
\colhead{Sub-sample} &
\colhead{(Completeness)} \\
}
\startdata
707/26 & $1.50\times 10^{-16}$  \tablenotemark{b} & 39.61  & 69	& 74.05~~~(93.2\%) \nl	
814/33 & $0.71\times 10^{-16}$  \tablenotemark{c} & 40.30  & 423	& 431.76~~(98.0\%) \nl
909/40 & $0.57\times 10^{-16}$  \tablenotemark{c} & 40.67  & 168	& 178.42~~(94.2\%) \nl
 &  &	 & 	\nl
Overall: &                      &        & 660	&  684.22~~(96.5\%) \nl
\enddata
\tablenotetext{a}{Limiting \Ha\ luminosity (\ergsPersec).}
\tablenotetext{b}{flux cut placed higher than that of the original homogeneous sub-sample}
\tablenotetext{c}{identical flux cut to that of the original homogeneous sub-sample;
{\em i.e.}~homogeneous sub-sample consists entirely of \Ha\ candidates.}
\end{deluxetable}

\clearpage
\begin{deluxetable}{lcccc}
\tablewidth{0pc}
\tablecaption{TTF Detection of Autofib Sub-Sample\label{t:figautofib}}
 
\tablehead{
\colhead{Range}	& 
\colhead{TTF} & 
\colhead{Undetected}	& 
\colhead{Line-Only} &
\colhead{Line-on-} \\
\colhead{ }	& 
\colhead{Detections\tablenotemark{a}} & 
\colhead{ }	& 
\colhead{ } &
\colhead{Continuum} \\
}
\startdata
 			&		&	  &	   &	  \nl
By $b_J$ magnitude:	&		&	  &	   &	  \nl
 			&		&	  &	   &	  \nl
$18.0 \leq b_J < 19.0$ &	$ 1/1 $ &	0 &	0 &	1 \nl
$19.0 \leq b_J < 20.0$ & 	$ 4/4 $ &	0 &	1 &	3 \nl
$20.0 \leq b_J < 21.0$ & 	$ 4/5 $ &	1 &	1 &	3 \nl
$21.0 \leq b_J < 22.0$ &	$ 3/7 $ &	4 &	2 &	1 \nl
$22.0 \leq b_J < 23.0$ & 	$ 1/1 $ &	0 &	1 &	0 \nl
 			&		&	  &	   &	  \nl
 			&		&	  &	   &	  \nl
By \OII\ EW:		&		&	  &	   &	  \nl
 			&		&	  &	   &	  \nl
${\rm EW} \geq 60$ \AA\	&	$ 1/2 $	&	1 &	0 &	1 \nl
$50 \leq {\rm EW} < 60$ \AA\ &	$ 0/1 $ &	1 &	0 &	0 \nl
$40 \leq {\rm EW} < 50$ \AA\ &	$ 1/1 $ &	0 &	0 &	1 \nl
$30 \leq {\rm EW} < 40$ \AA\ &	$ 5/6 $ &	1 &	1 &	4 \nl
$20 \leq {\rm EW} < 30$ \AA\ &	$ 1/2 $ &	1 &	0 &	1 \nl
$10 \leq {\rm EW} < 20$ \AA\ &	$ 5/6 $ &	1 &	4 &	1 \nl
\enddata
\tablenotetext{a}{number detected by TTF out of the number of Autofib galaxies
available}
\end{deluxetable}

\clearpage
\begin{deluxetable}{lcccc}
\tablewidth{0pc}
\tablecaption{Emission-Line Occurrence in Homogeneous 
Sub-Sample\label{t:emisstype}}

\tablehead{
\colhead{Filter $\lambda/\Delta \lambda$}	& 
\colhead{Line-Only}	& 
\colhead{Line-Only} &
\colhead{Line-on-Continuum} & 
\colhead{Line-on-Continuum}	\\
\colhead{(nm)} & 
\colhead{1 Slice}	& 
\colhead{2 Slices} &
\colhead{1 Slice} & 
\colhead{2 Slices}	\\
}
\startdata
707/26		& 75		& 0		& 15		& 15 \nl
814/33		& 381		& 6		& 23		& 13 \nl
909/40		& 138		& 3		& 14		& 13 \nl
               &               &               &                &    \nl
Total:		& 594		& 9		& 52		& 41 \nl
\enddata
\end{deluxetable}


\clearpage 

\figcaption[Distribution of fields on the sky]
{Distribution of survey regions on the sky.
Regions are labeled with the number of fields contained in each.
Control fields ({\em triangles}) are randomly chosen high galactic
latitude regions. The remaining fields ({\em circles}) were deliberately
chosen to overlap with Autofib Survey galaxies. The north and south 
galactic poles (NGP/SGP) are also indicated ({\em crosses}).
\label{figaitoff}}

\figcaption[Evolution of \cite{kennicutt92a}~(1992a) line ratios with redshift]
{Evolution of the mean \cite{kennicutt92a}~(1992a) line-flux ratios with
redshift. All ratios are expressed relative to (\Ha$+$\NII) ({\em solid
lines}) except \NII\ which is shown relative to \Ha\ alone 
({\em dashed line}). 
Change in the ratios is due to the weighting of the \cite{kennicutt92a}
sample by the numbers of galaxy type found at each redshift ({\em inset}).
Galaxy densities are taken from the evolution of $\phi^*(z)$ for individual
galaxy types as given by \cite{heyl97}~(1997).
\label{figkenevol}}

\figcaption[Predicted occurrence of emission-lines in the TTF sample]
{Predicted relative occurrence of emission-line galaxies in each 
of the three wavelength regions of the {\sl TTF Field Galaxy
Survey}. Shown are the relative numbers per flux bin ({\em left}) and
the cumulative numbers to fainter limiting flux ({\em right}).
\label{figcontamEVOL}}

\figcaption[Example scans and TTF spectra of emission-line candidates]
{({\em top}) Example strip-mosaic scans of a subset of candidates
from a {\sl TTF Field Galaxy Survey} field. Individual images
are 9\as\ on a side with north at top, east to the left. Circles denote
aperture size.
({\em bottom}) TTF spectra for the same galaxies. Preliminary 
({\em dotted line}) and final ({\em solid line}) continuum fits. Numbers
shown ({\em right}) are flux ($\times 10^{-16}$ \ergs2band), 
star-galaxy classification parameter and $\sigma$-deviation.
Deviant points (excluded from the final continuum fit) are also
indicated ({\em circles}). The zero flux level is shown by the
horizontal tickmarks (where present) and non-detections are
represented on this level ({\em crosses}). Galaxy 214.14
is one of the 18 Autofib objects in common with our fields.
All of these objects are taken from a field observed in the 707/26 interval.
\label{d13e7Best}}

\figcaption[Distribution of $\sigma$-deviations throughout the sample]
{($a$) Distribution of $\sigma$-deviations for the preliminary catalogue of
emission-line candidates selected with $\sigma$-deviation $\geq 3$.
Line-only ({\em horizontal tickmarks}) and line-on-continuum 
detections ({\em crosses})
are shown separately and the fields are grouped according to
blocking filter spectral locations (707, 814 or 909~nm).
($b$) Distributions of $\sigma$-deviations across all fields
pertaining to the 707 ({\em solid}), 814 ({\em dotted}) and 909 
({\em dashed}) filters.
($c$) Cumulative distributions of those shown in ($b$).
\label{figlimits_sig}}

\figcaption[Line-only candidates superposed on detection probability contours]
{({\em left}) Line-only candidates shown in terms of both mean
detected (emission-line) flux and object size 
(full-width at half-maximum, FWHM), for each of the
707/26, 814/33 and 909/40 emission-line samples. 
The 22\_3A (707) and 10\_3C (909) candidates have been removed.
The detection probability
contours ({\em solid lines}) indicate levels of 20, 40, 60 and 80\%
and are taken from the {\sl single least-sensitive} field in each group.
All other fields have contours at fainter limits (not shown).
The 60\% contour is highlighted ({\em bold line}).
({\em right}) Same sample of line-only candidates with the chosen
cuts of line flux and surface brightness overlaid. Members of the 
homogeneous sub-sample lie above these two lines.
\label{figlimits5}}

\figcaption[Line-on-continuum candidates superposed on detection probability
contours]
{({\em left}) Line-on-continuum candidates shown in terms of both mean
detected (continuum) flux and object size 
(full-width at half-maximum, FWHM), for each of the
707/26, 814/33 and 909/40 emission-line samples. 
The 22\_3A (707) and 10\_3C (909) candidates have been removed.
Identical flux and surface brightness cuts to those applied to the line-only
sample in Fig.~\ref{figlimits5} are also shown ({\em solid lines}).
({\em right}) Same sample of line-on-continuum objects showing the
scatter of continuum flux and emission-line flux. The flux cuts
applied at {\em left} have also been applied here to both the
line and continuum ({\em vertical and horizontal lines}).
The detection probability
contours ({\em solid lines}) indicate levels of 20, 40, 60 and 80\%
and are taken from the {\sl single least-sensitive} field in each group.
All other fields have contours at fainter limits (not shown).
The 60\% contour is highlighted ({\em bold line}).
Lines showing the cut in observed equivalent width are shown labeled
with their respective values ({\em solid diagonal line}).
The {\em dotted line} shows the limiting locus for 3$\sigma$-deviation
objects as measured through the largest aperture.
\label{figlimits6}}

\figcaption[Completeness as a function of emission-line flux]
{Cumulative completeness of the homogeneous sub-samples in the 
($a$) 707/26, ($b$) 814/33 and ($c$) 909/40 filters, 
in the limit of fainter flux.
\label{figcomplete2}}

\figcaption[Comparison between Autofib and TTF measurements]
{Comparison between Autofib and TTF flux measurements, with the
measurements from each plotted on the abscissa and ordinate
of each panel. ($a$) TTF-measured (\Ha$+$\NII)
line fluxes against Autofib \OII\ equivalent widths. 
($b$) TTF-measured (\Ha$+$\NII)
equivalent widths against those of \OII\ from Autofib.
The row of points along the top denote galaxies detected by TTF
for which the equivalent width was indeterminable.
($c$) TTF-measured continuum flux versus $b_J$ photographic magnitude.
The best-fit $b_J = -2.5 \log F({\rm cont}) + 21.81$ is also shown
({\em dotted line}), weighted by the $1/\Delta F^2$ continuum
photometry uncertainties.
\label{figautofib}}

\figcaption[Redshift distribution of H$\alpha$ emission-line 
candidates by field]
{Redshift distributions as measured from the subset of \Ha\
emission-line candidates. Distributions shown for each field
in the ($a$) 707/26, ($b$) 814/33, and ($c$) 909/40 sets.
\label{figpeaks2}}

\figcaption[Narrowband number counts for the homogeneous sub-sample of all
emission-line candidates]
{Narrowband number counts for the homogeneous sub-sample
of emission-line candidates above the 
4 ({\em filled circles}) and $5\sigma$ ({\em open circles}) limits
of survey selection.
Shown are cumulative numbers in each of the 707/26, 814/33 and 909/40 sets. 
All counts are expressed in terms
of the total object number per flux bin, per square degree of sky,
per 100~\AA\ of scan interval. Counts have been corrected through the
incompleteness weights derived in Sect.~\ref{s:defHomog}. 
Flux bins are $0.1 \log [F({\rm H}\alpha)]$ and the
flux limits for each of the \Ha\ sub-samples (forming only {\sl part} of 
these sets) are also shown ({\em vertical dotted lines}).
The {\em solid line} shows evolution constrained by 
the \cite{gallego95}~(1995) and \cite{tresmad98}~(1998) \Ha\
luminosity functions, as detailed in Sect.~\ref{s:volLimited}.
\label{fignumbersEVOL2}}

\figcaption[Preliminary \Ha\ luminosity functions]
{Preliminary \Ha\ luminosity functions 
from the {\sl TTF Field Galaxy Survey} ({\em filled circles})
at mean redshifts of 
($a$) $\bar{z} = 0.08$, ($b$) $\bar{z} = 0.24$, and 
($c$) $\bar{z} = 0.4$. 
No extinction corrections have been applied to the luminosities.
The other points are the \Ha\ luminosity functions from
\cite{gallego95}~(1995; {\em open triangles}) for $z \simlt 0.045$
and \cite{tresmad98}~(1998; {\em open circles}) at $z \sim 0.2$.
\label{figLFdn}}


\clearpage
\centerline{\epsfxsize=140mm\epsfbox{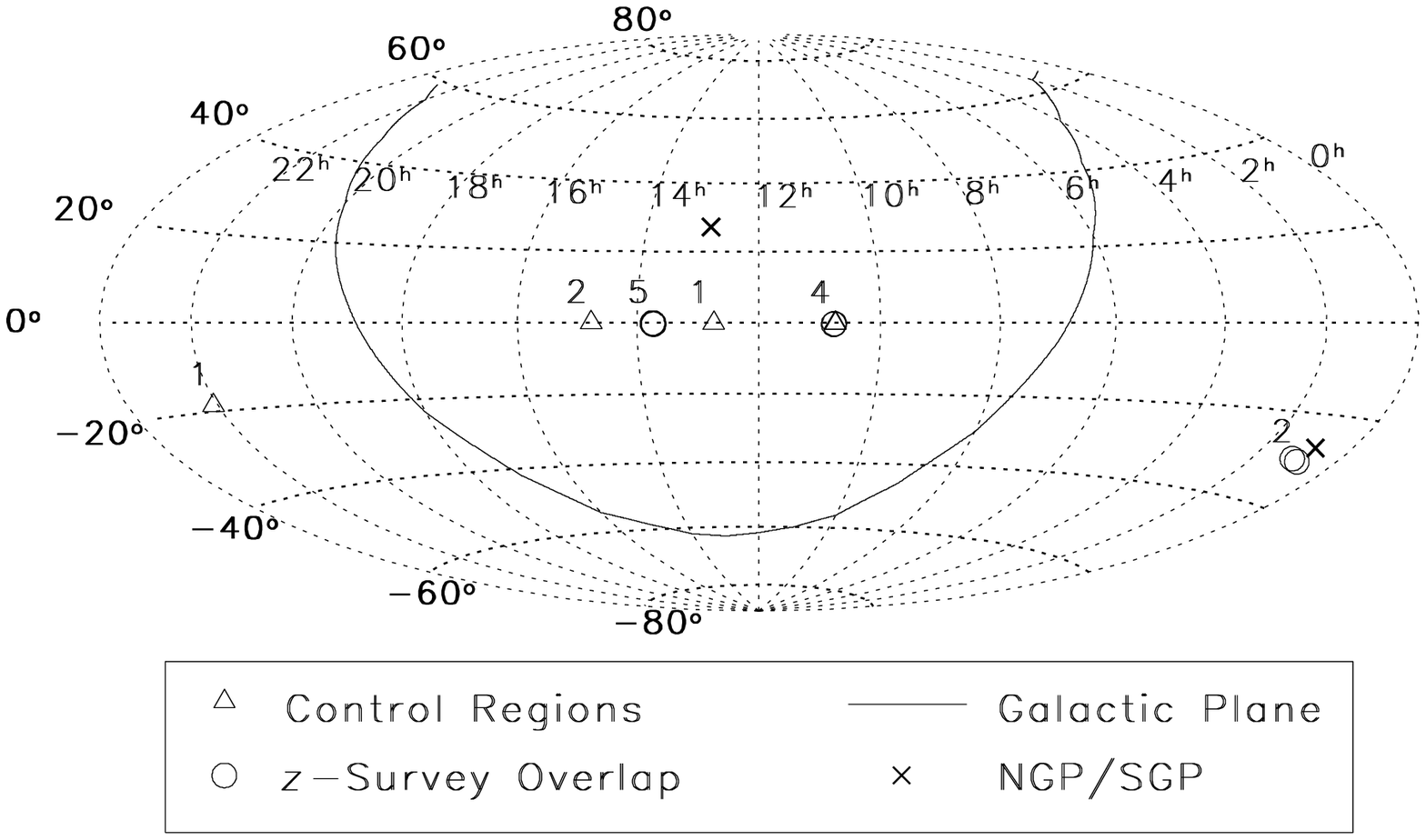}}
\vspace{1cm}
{\bf \Large \hspace{7cm} Figure 1}

\clearpage
\centerline{\epsfxsize=130mm\epsfbox{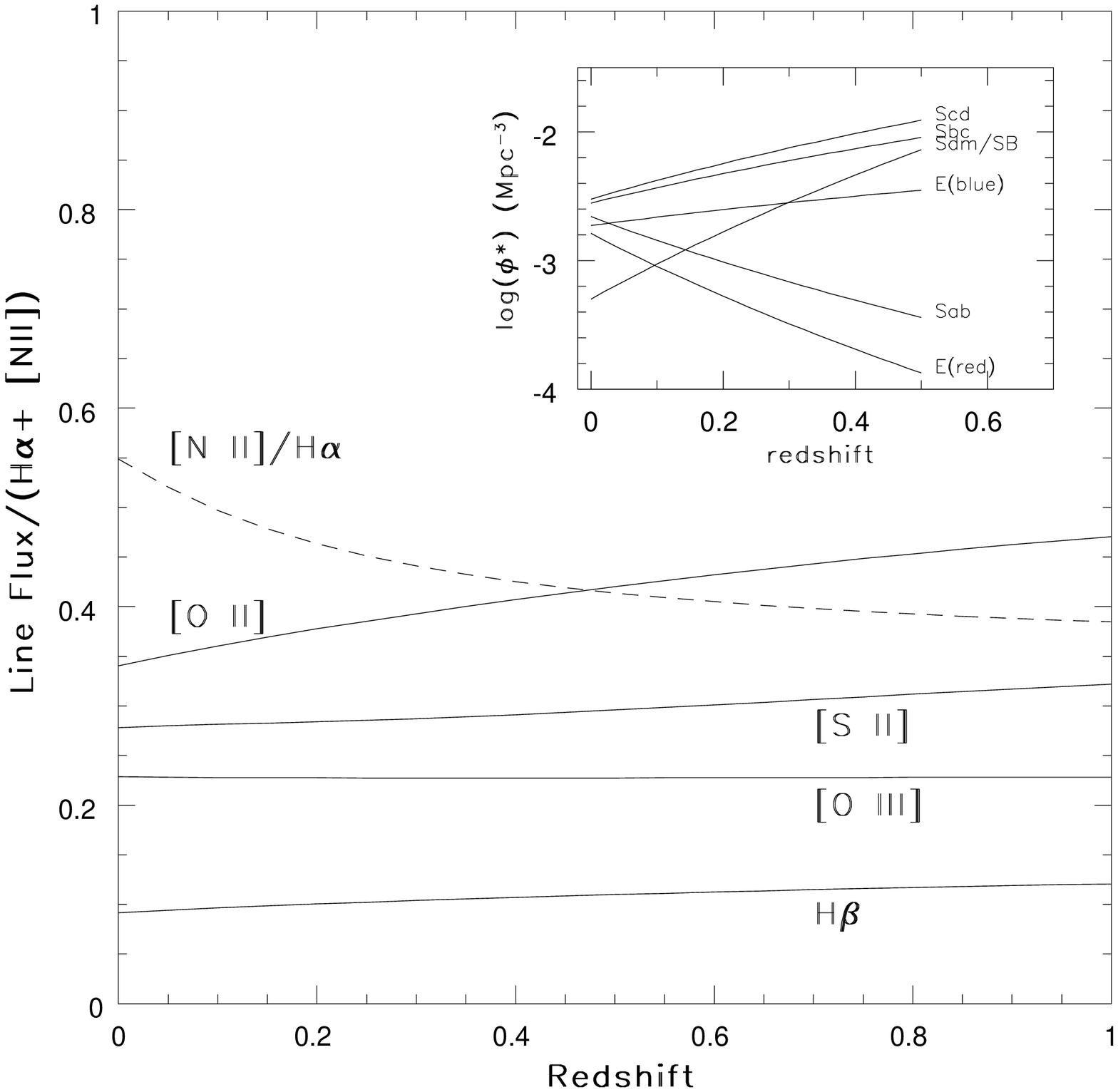}}
\vspace{1cm}
{\bf \Large \hspace{7cm} Figure 2}

\clearpage
\centerline{\epsfxsize=150mm\epsfbox{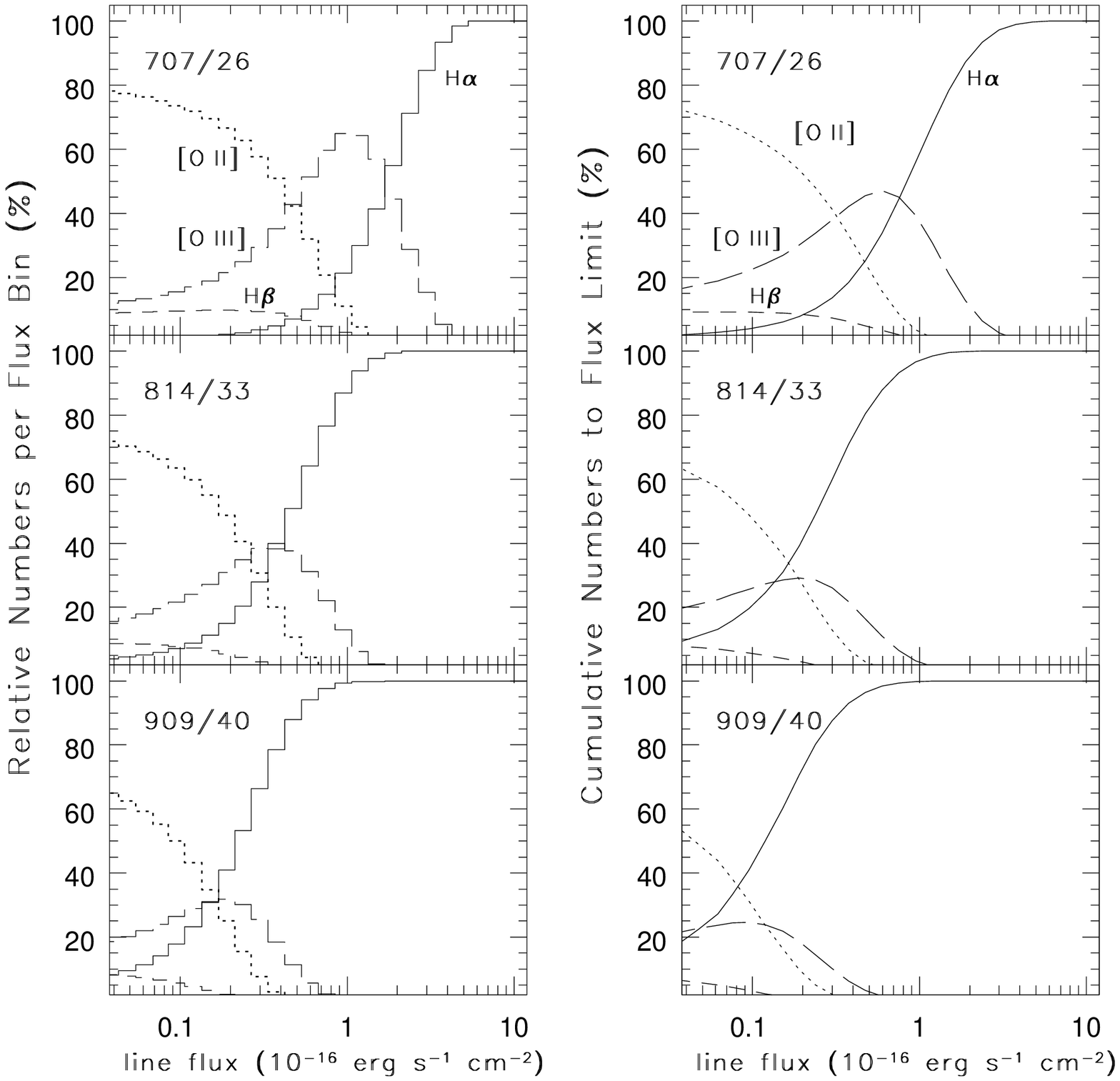}}
\vspace{1cm}
{\bf \Large \hspace{7cm} Figure 3}

\clearpage
\centerline{\epsfxsize=140mm\epsfbox{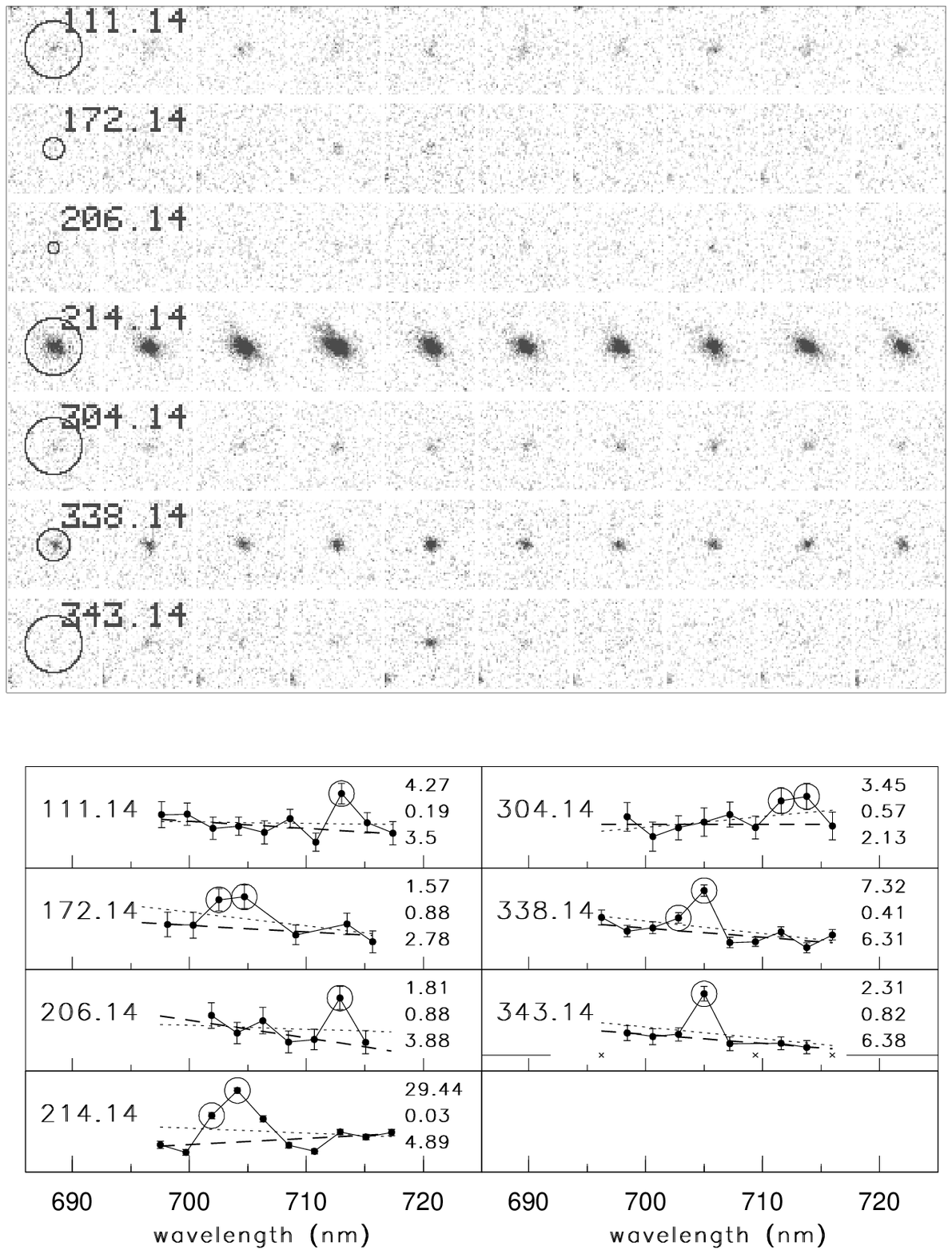}}
\vspace{1cm}
{\bf \Large \hspace{7cm} Figure 4} 

\clearpage
\centerline{\epsfxsize=140mm\epsfbox{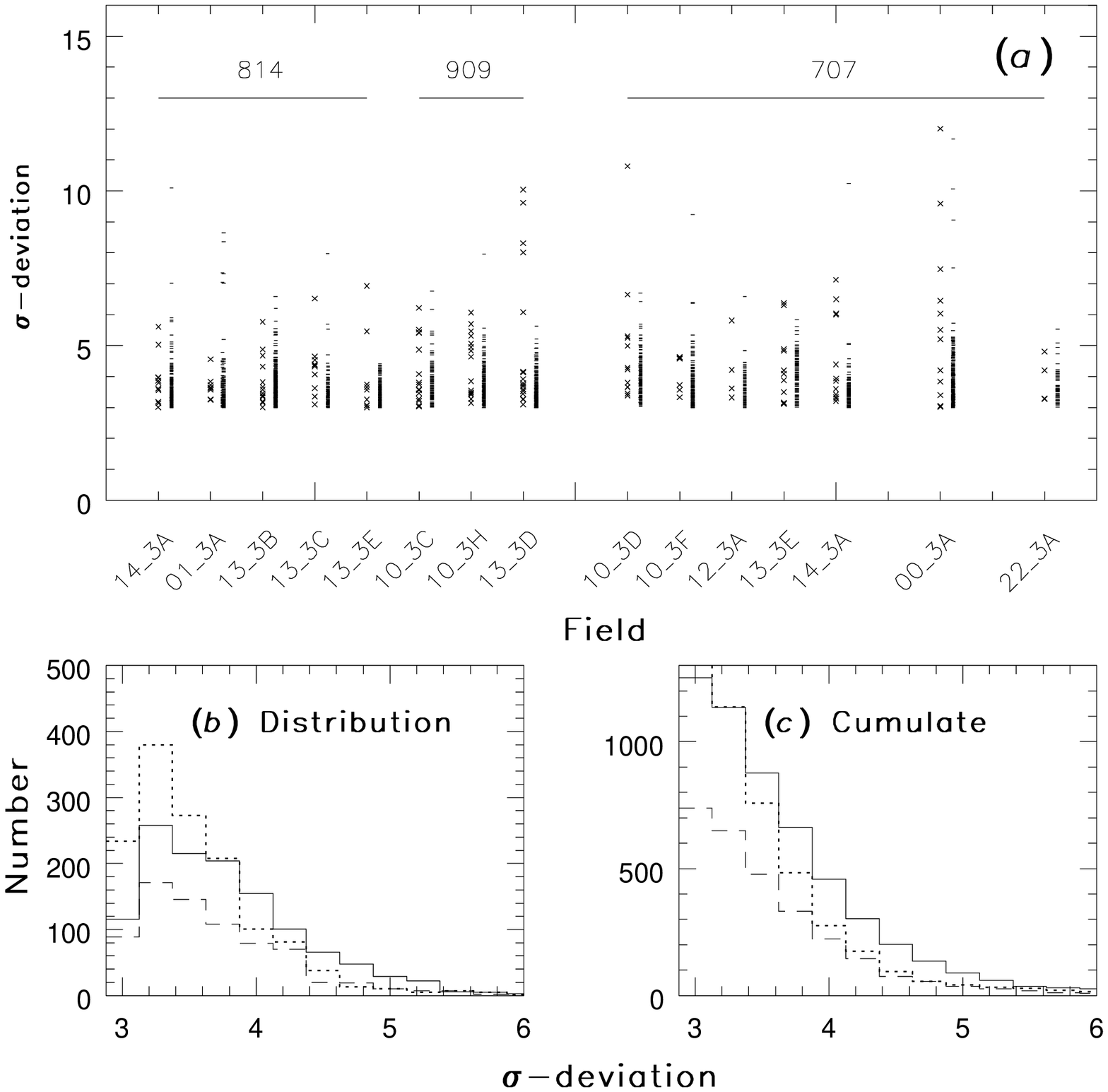}}
\vspace{1cm}
{\bf \Large \hspace{7cm} Figure 5}

\clearpage
\centerline{\epsfxsize=150mm\epsfbox{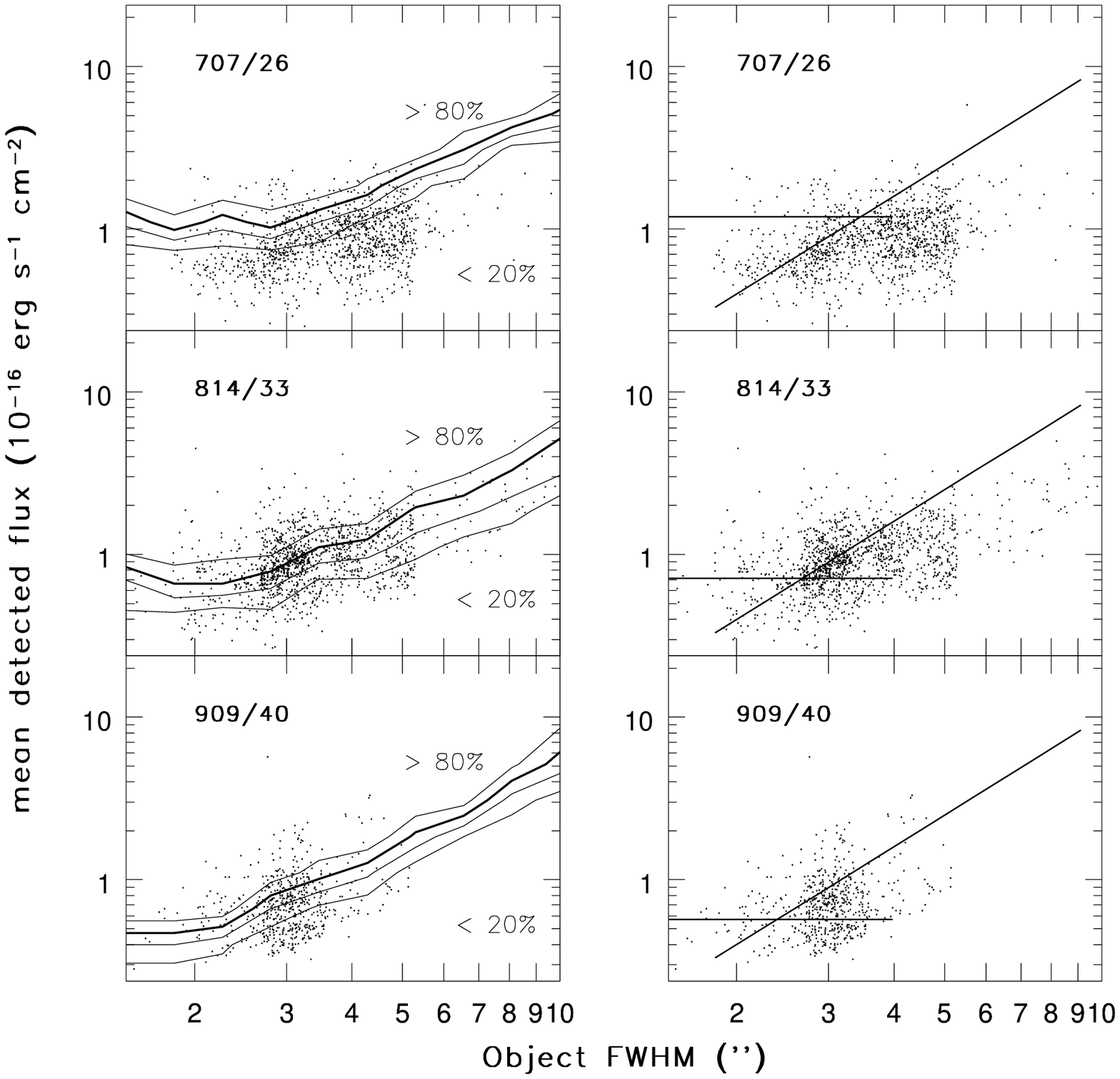}}
\vspace{1cm}
{\bf \Large \hspace{7cm} Figure 6} 

\clearpage
\centerline{\epsfxsize=150mm\epsfbox{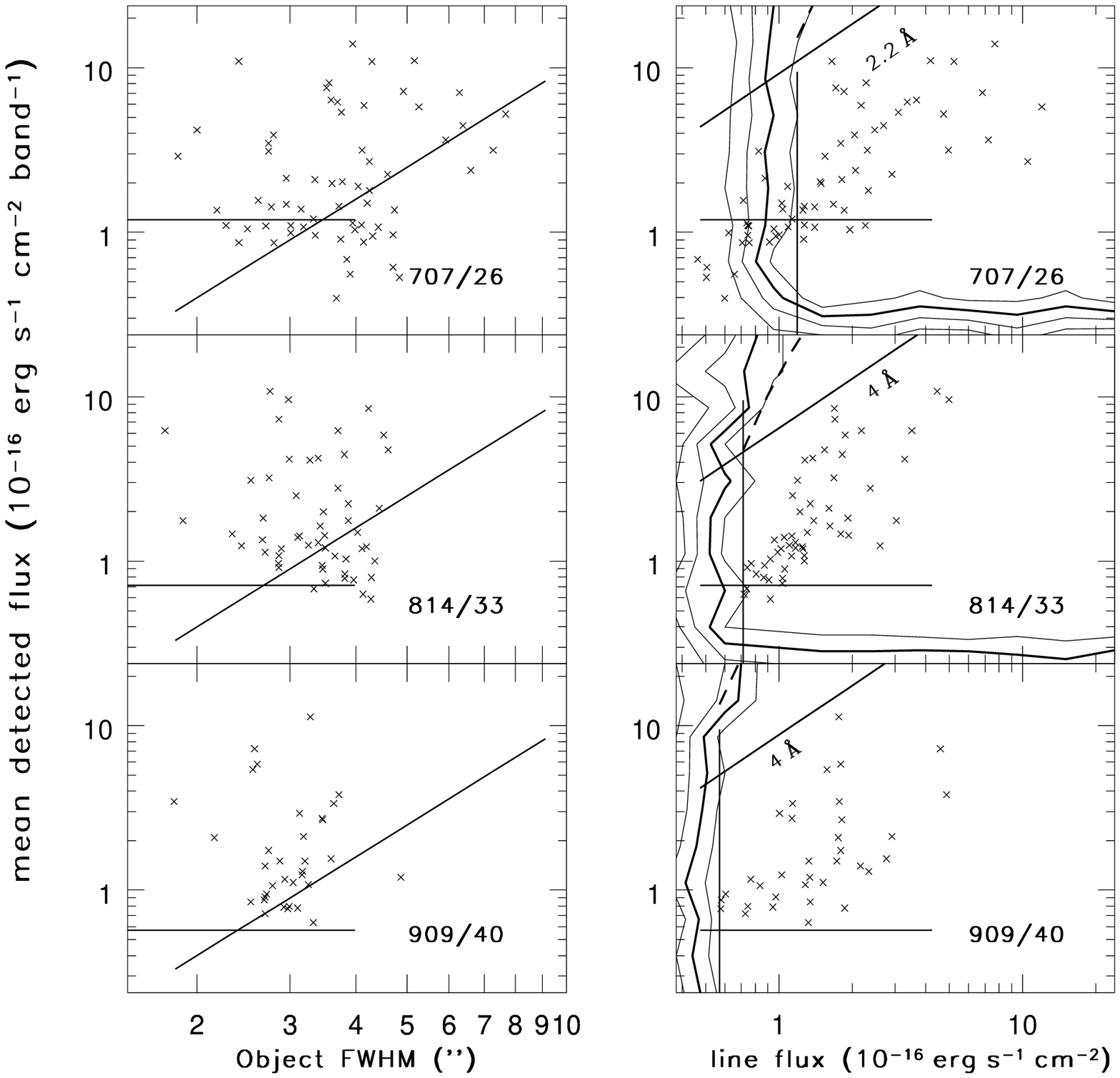}}
\vspace{1cm}
{\bf \Large \hspace{7cm} Figure 7} 

\clearpage
\centerline{\epsfxsize=110mm\epsfbox{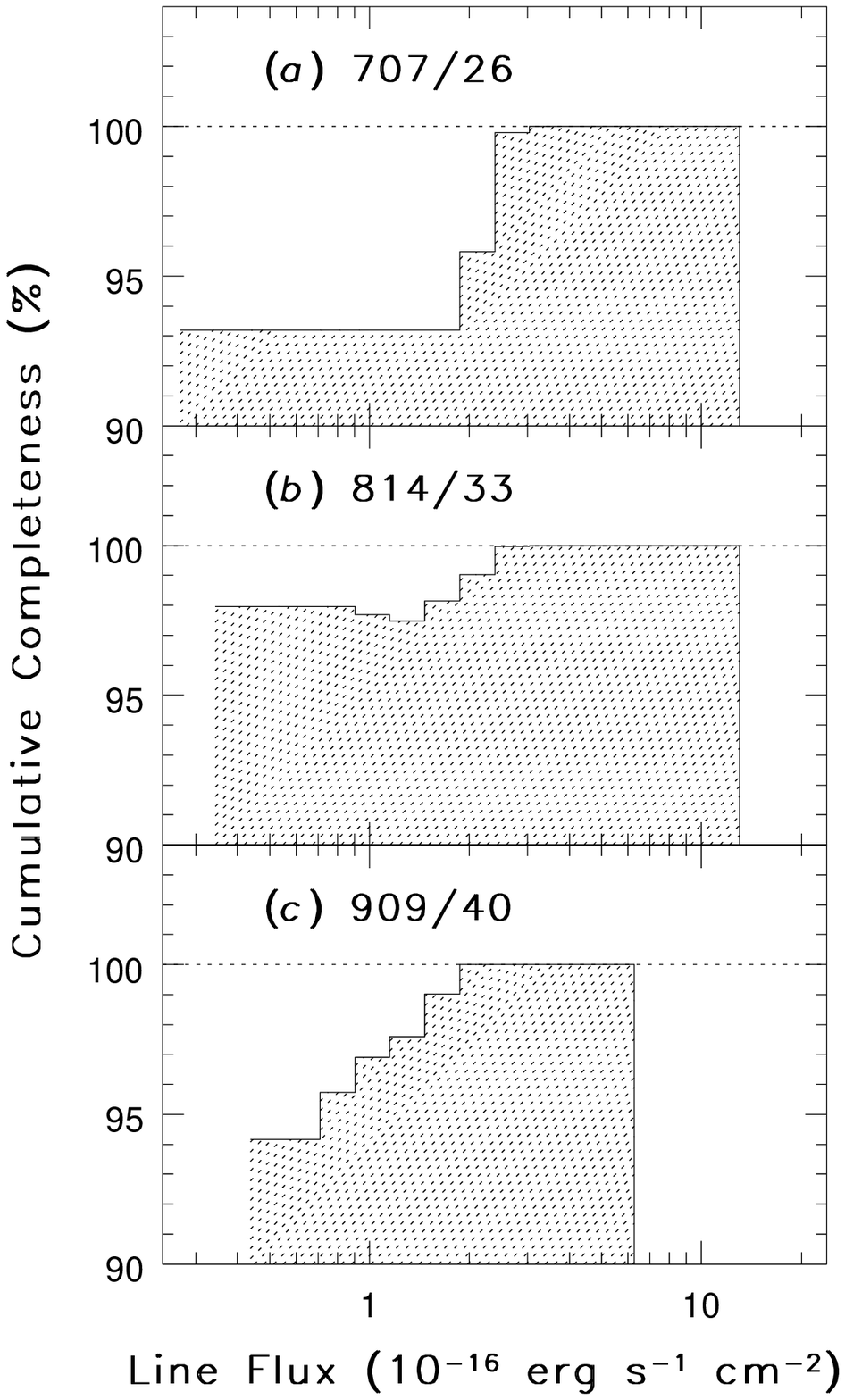}}
\vspace{1cm}
{\bf \Large \hspace{7cm} Figure 8}

\clearpage
\centerline{\epsfxsize=145mm\epsfbox{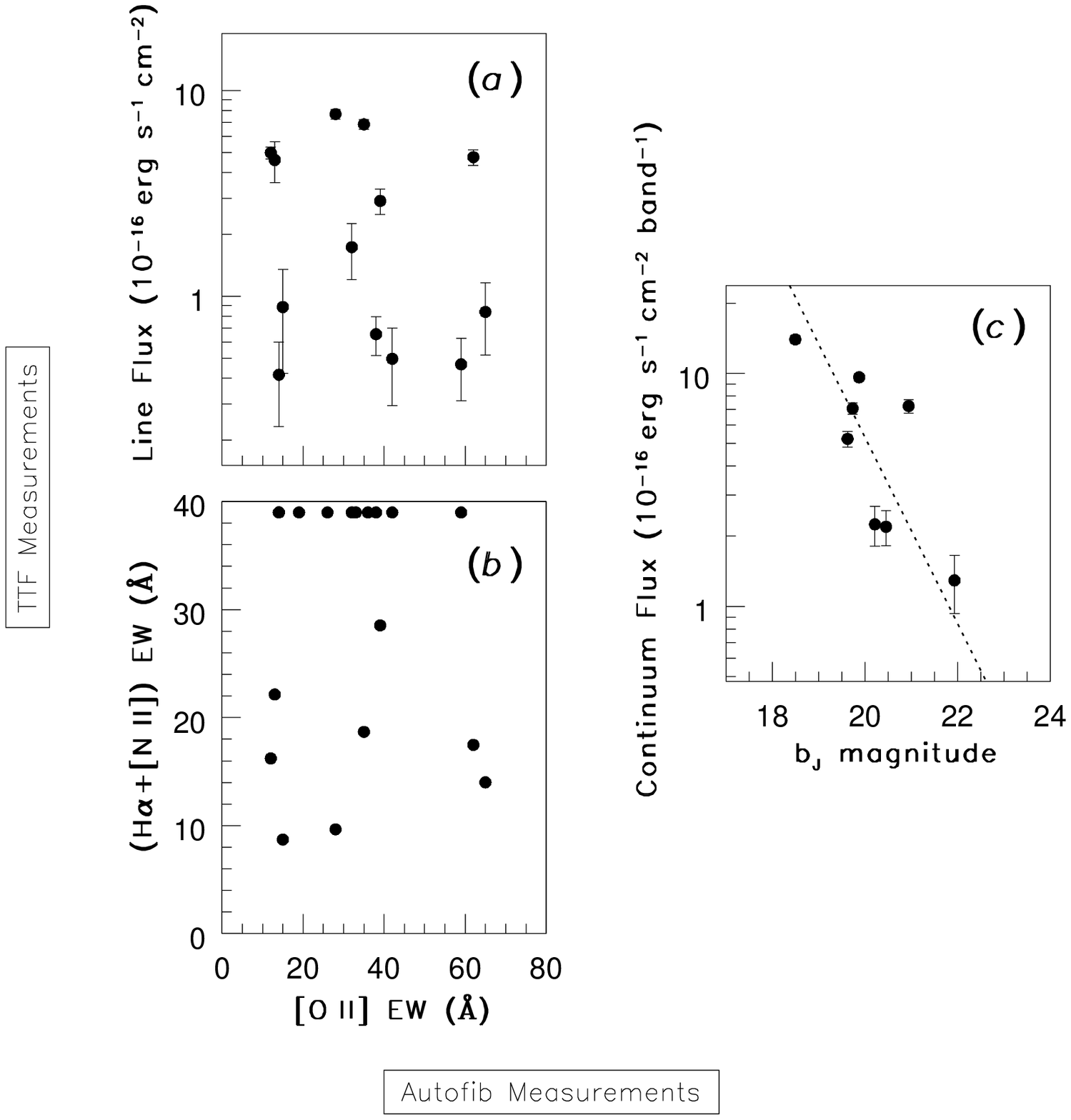}}
\vspace{1cm}
{\bf \Large \hspace{7cm} Figure 9}

\clearpage
\centerline{\epsfxsize=140mm\epsfbox{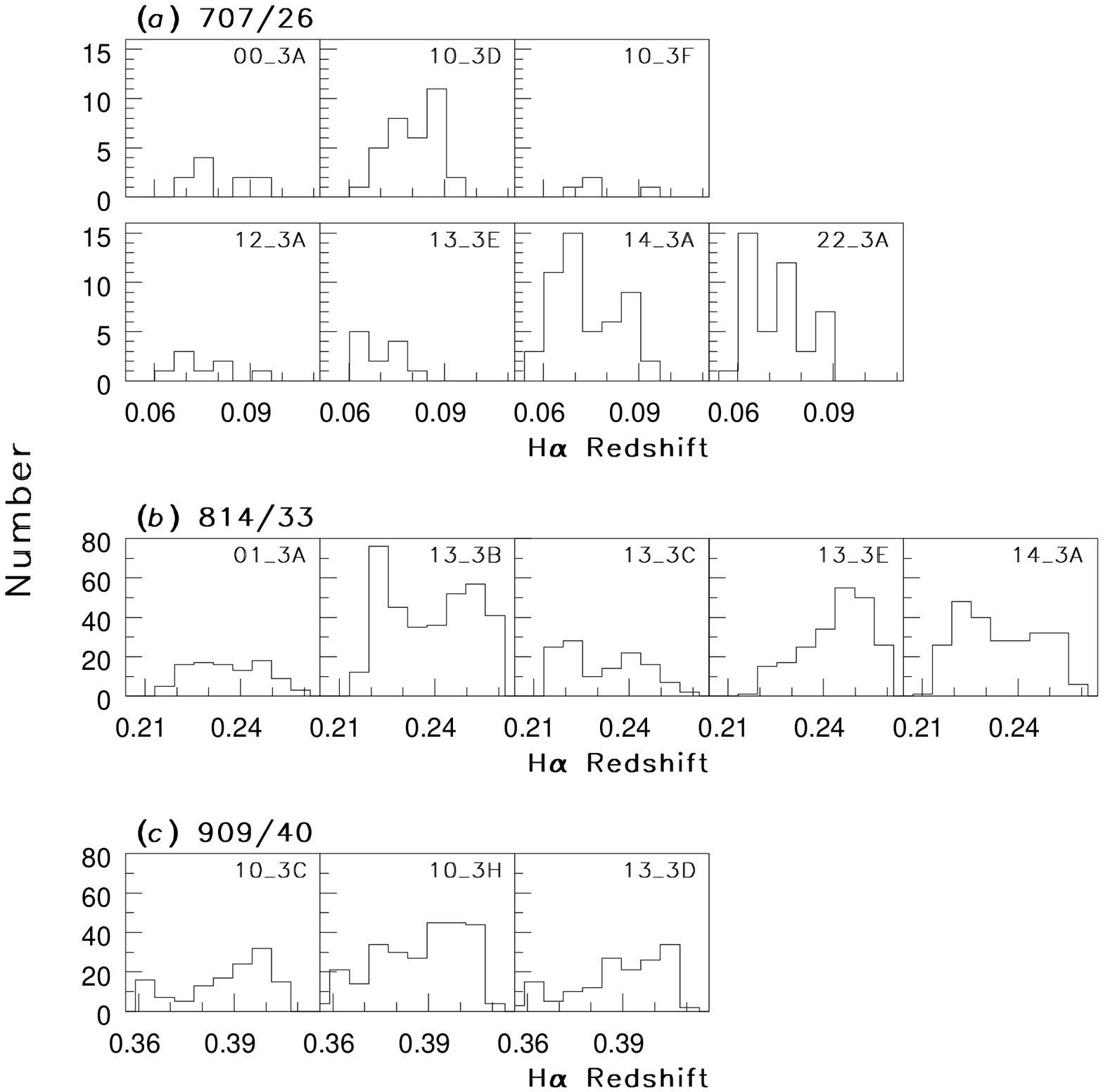}}
\vspace{1cm}
{\bf \Large \hspace{7cm} Figure 10}

\clearpage
\centerline{\epsfxsize=120mm\epsfbox{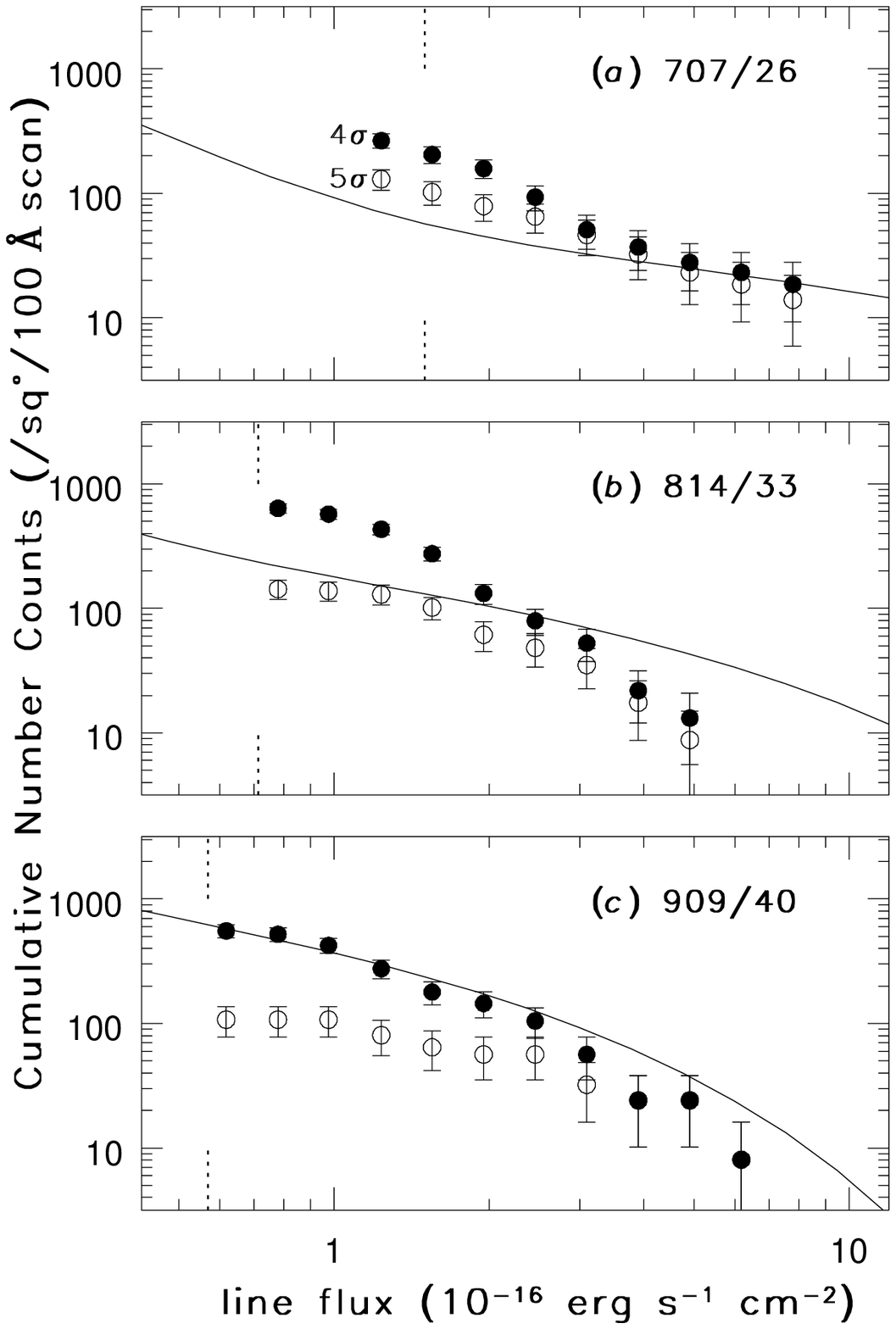}}
\vspace{1cm}
{\bf \Large \hspace{7cm} Figure 11}

\clearpage
\centerline{\epsfxsize=130mm\epsfbox{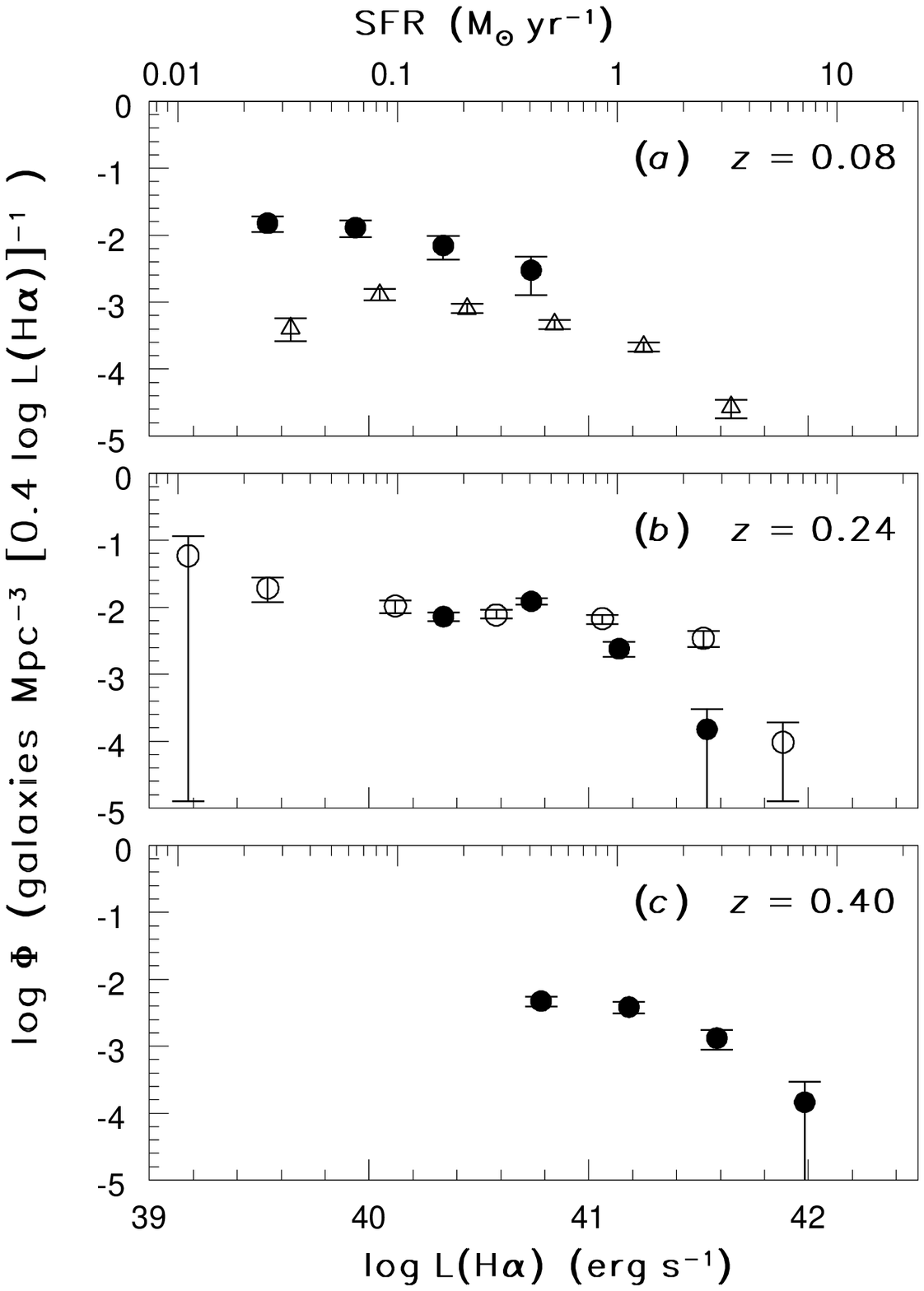}}
\vspace{1cm}
{\bf \Large \hspace{7cm} Figure 12} 

\end{document}